\documentclass[aip,pof,reprint]{revtex4-2}
\usepackage{amsmath,amssymb,graphicx}
\usepackage{pgfplots}
\usepackage{tikz}
\usetikzlibrary{plotmarks,positioning,shapes,arrows,backgrounds}
\pgfplotsset{compat = newest}
\pgfplotsset{legend style={font=\footnotesize}}
\pgfplotsset{samples=200}

\begin{document}

\title{Semiclassical Structure of the Advection--Diffusion Spectrum in Mixed Phase Spaces}

\author{Christopher Amey}
\affiliation{Department of Physics, Brandeis University,
Waltham, Massachusetts 02453}
\author{Bala Sundaram}
\affiliation{Department of Physics, University of Massachusetts,
100 Morrissey Boulevard, Boston, Massachusetts 02125}
\author{Andrew C. Poje}
\affiliation{Graduate Faculty in Physics \& Department of Mathematics,
City University of New York - CSI, Staten Island, New York 10314}
\date{\today}

\begin{abstract}
We examine the spectral structure of the two-dimensional advection–diffusion operator in flows with mixed phase space at very large Péclet number. Using Fourier discretization combined with symmetry reduction and Krylov–Arnoldi methods, we compute on the order of one hundred leading eigenpairs reliably in the asymptotic, weak-diffusion regime. While the principal eigenvalue is asymptotically diffusive and localized on the largest regular region, the broader spectrum exhibits a rich organization controlled by local Lagrangian phase-space geometry. In particular, exponential mixing in chaotic regions rapidly suppresses correlations, whereas algebraic mixing in integrable regions generates long-lived coherent structures that dominate the slow and intermediate parts of the spectrum. We identify three distinct classes of eigenmodes: advective modes associated with transport on invariant tori, diffusive modes and, within the duffusive branch, tunneling modes arising from weak coupling between dynamically separated regular regions. Drawing on a semiclassical analogy, we assign quantum-number-like labels to these families and predict the appearance, scaling, and ordering of their sub-spectra directly from the Hamiltonian phase-space structure. The coexistence of these families implies that no uniform control of the spectral gap exists across the full spectrum: although the slowest mode is diffusive, arbitrarily small gaps arise between competing families at higher mode numbers. As a result, finite-time advection–diffusion dynamics is generically governed by persistent modal competition rather than single-mode dominance, even at asymptotically large Péclet number.
\end{abstract}

\maketitle
\section{Introduction}

The efficient mixing of passive scalars plays a fundamental role in settings ranging from
industrial processing and microfluidic design to environmental and geophysical transport.
In the absence of turbulence, such mixing is often achieved via \emph{chaotic advection},
where exponential stretching and repeated folding of material lines create fine-scale
gradients on which molecular diffusion can act \cite{Aref2017,Babiano1994}. These mechanisms
underlie a broad class of transport processes extending over many orders of magnitude in
physical scale, from micrometer-scale microfluidic mixers \cite{Benzekri2006,Schlick2013} to
planetary atmospheres and oceanic flows where coherent vortices and jets inhibit tracer
dispersion and lead to anomalous transport \cite{Lopez2001,Prants2011}.

Realistic time-dependent flows exhibit a wide range of dynamical behavior depending on 
their kinematic structure. At one extreme lie globally integrable flows, where trajectories 
are confined to invariant tori and material elements experience only algebraic 
deformation in time. In this setting the advection--diffusion operator decomposes 
naturally into an axisymmetric sector, which is self-adjoint and purely diffusive 
with decay rates $\gamma \sim Dk^2$ (or equivalently, $\gamma \sim \mathrm{Pe}^{-1}$), 
and non-axisymmetric sectors that remain intrinsically non-self-adjoint. Rigorous 
semiclassical analysis shows that these advective sectors support families of modes 
whose decay rates scale anomalously as $\mathrm{Pe}^{-1/2}$ or $\mathrm{Pe}^{-1/3}$, 
depending on the local structure of the frequency map.

At the opposite extreme lie globally chaotic flows. For uniformly hyperbolic systems, 
rigorous analysis establishes logarithmic enhanced dissipation with decay rates 
$\gamma \sim \log(\mathrm{Pe})$ \cite{elgindi2025optimal}. More generally, 
in non-uniformly hyperbolic chaotic systems, strong spatial and temporal variability 
in stretching rates produces rich spectral structure including strange eigenmodes 
\cite{Pierrehumbert:1994} and multifractal scalar measures 
\cite{Toussaint2000,Sundaram:2009PRE,Popovych2007}. In such flows, the spectrum of 
the one-period advection--diffusion operator reflects the distribution of finite-time 
Lyapunov exponents and the history of filamentation 
\cite{Mezic2005,Mezic2013,Froyland2010}.

Between these extremes lie \emph{mixed phase spaces}, in which chaotic regions 
coexist with dynamically invariant regular islands surrounding elliptic fixed points 
or invariant tori. These islands support algebraic deformation and long-lived coherent 
structures, introducing a second, qualitatively distinct transport mechanism within 
the same operator. Mixed phase spaces therefore combine three competing mechanisms 
within a single dynamical system: non-uniformly hyperbolic chaos, algebraic transport 
on regular sets, and diffusion across their boundaries. How these mechanisms jointly 
shape the spectrum of the advection--diffusion operator---beyond the leading 
eigenvalue---remains a central unresolved problem.

Significant progress has been made toward understanding aspects of this competition.
Coarse-grained transfer-operator and mapping-matrix approaches have demonstrated that, at
very large effective Péclet numbers, dominant decay modes localize on regular islands in
flows with strongly chaotic components
\cite{Giona2004,Cerbelli:2004,Gorodetskyi:2012PoFa,Gorodetskyi:2012PoFb,Schlick2013}. 
Gorodetskyi et al.~demonstrated that this localization persists even when regular regions 
occupy arbitrarily small fractions of phase space, provided the diffusive length scale 
remains smaller than the characteristic island width. These computational methods establish 
that the asymptotic decay in mixed phase spaces is controlled by island-localized, diffusive 
modes.

Popovych et al.~\cite{Popovych2007} observed that in certain parameter regimes 
where multiple eigenvalues are nearly degenerate, the resulting modal competition produces 
non-exponential decay over finite times, with eigenmodes localizing on elliptic islands or 
boundaries between weakly connected chaotic regions. However, they found the occurrence 
of such near-degeneracies to vary unpredictably in the 2D parameter space of their model.

Complementing these computational results, rigorous semiclassical analysis of integrable 
advection-diffusion operators has established the theoretical foundation for anomalous 
scaling in regular flows \cite{Vukadinovic2015,bedrossian2017,Vukadinovic2021}. These 
results provide sharp mathematical proofs that advective sectors in integrable flows 
support families of modes with $\mathrm{Pe}^{-1/2}$ or $\mathrm{Pe}^{-1/3}$ decay 
rates—the so-called advection-enhanced diffusion regime—with the specific exponent 
determined by local geometric structure near elliptic fixed points. However, these 
analyses rely on global action-angle coordinates and do not extend to mixed phase spaces 
where multiple island chains coexist with chaotic regions.

Extending these insights to systems with multiple coexisting regular regions of 
different scales presents both computational and conceptual challenges. Computational 
methods have established asymptotic localization and observed near-degeneracies in 
specific cases, while rigorous analysis provides scaling laws for globally 
integrable flows. What remains unexplained is how the full spectrum organizes when 
several island chains compete: which modes dominate finite-time dynamics, when spectral 
gaps collapse, and in particular, when and why competing island families produce 
near-degenerate eigenvalues. Addressing these questions requires an interpretive 
framework capable of predicting spectral organization from the underlying phase-space 
geometry.

The advection-diffusion operator admits a semiclassical interpretation. In this analogy, 
the diffusivity $D$ plays the role of an effective Planck constant, so that the limit 
$\mathrm{Pe}\to\infty$ corresponds to $\hbar_{\mathrm{eff}}\to 0$. Regular islands act as 
finite potential wells supporting diffusive modes with $\mathrm{Pe}^{-1}$ scaling, while 
elliptic cores behave as harmonic oscillators producing advective mode families with 
$\mathrm{Pe}^{-1/2}$ spacing. The chaotic sea mediates weak coupling between these 
families, giving rise to tunneling-like hybridization when wells approach degeneracy. This 
interpretation identifies geometric quantities—island size and core curvature—as the 
fundamental parameters controlling spectral organization, and provides a predictive 
framework for when distinct spectral branches arise and how they scale.

In this work, we compute the leading eigenvalues and eigenmodes of the one-period 
advection-diffusion operator for the Chirikov standard map at Péclet numbers up to $10^7$. 
We show that the spectrum organizes into universal families—diffusive modes localized on 
islands, advective modes concentrated at elliptic cores, and hybrid modes spanning 
multiple regions—whose multiplicity and ordering can be predicted directly from 
phase-space geometry. When diffusive ladders associated with distinct islands approach one 
another, avoided crossings produce mixed states exhibiting the characteristic splittings 
of semiclassical tunneling. Importantly, although the slowest mode is ultimately diffusive 
and island-localized, the coexistence and hybridization of these families precludes 
uniform control of spectral gaps: arbitrarily small gaps arise between competing families 
at higher mode numbers, and finite-time dynamics is generically governed by persistent 
modal competition rather than single-mode dominance.

These results demonstrate that the local Hamiltonian geometry of the Lagrangian phase 
space determines the organization of the advection--diffusion spectrum. 
The semiclassical 
framework developed here provides both a classification scheme for eigenmodes in mixed 
systems and quantitative predictions for their appearance, validated by direct computation 
at asymptotically large Péclet numbers.

\section{Model and Spectral Formulation}

We consider the initial--value problem for the advection--diffusion equation on 
the two-torus $\mathbb{T}^2 = [0, 2\pi)^2$,
\begin{equation}
\frac{\partial\rho}{\partial t} + \vec{u}(x,t) \cdot \nabla\rho = D\nabla^2\rho, 
\quad x \in \mathbb{T}^2, \quad t > 0,
\label{eq:advdiff}
\end{equation}
with prescribed initial condition
\begin{equation}
\rho(x,0) = \rho_0(x), \quad x \in \mathbb{T}^2,
\end{equation}
and periodic boundary conditions in both spatial directions. Here $\rho(x,t)$ is 
a passive scalar, $D > 0$ is the molecular diffusivity, and $\vec{u}(x,t)$ is a 
given, divergence-free, time-periodic velocity field with period $T$.

Throughout, $\vec{u}$ is generated by the Chirikov--Taylor standard map \cite{Chirikov1979}. This 
provides a canonical setting in which integrable, mixed, and strongly chaotic 
phase-space structures can coexist within a single model, allowing us to probe 
how the spectral properties of the advection--diffusion operator reflect the 
underlying Hamiltonian geometry of the flow. The map arises from the time-periodic 
stream function
\begin{equation}
\psi(x, y, t) = \frac{y^2}{2} + K \cos x \sum_{n=-\infty}^{\infty} \delta(t - nT),
\end{equation}
where $\delta(t)$ is the Dirac delta function and the parameter $K$ controls the 
dynamics: $K = 0$ is integrable, intermediate values yield mixed phase spaces 
containing both chaotic seas and regular islands, and $K \to \infty$ approaches 
the anti-integrable limit with globally chaotic dynamics.

\begin{figure*}

\centering
{\includegraphics[width=\textwidth]{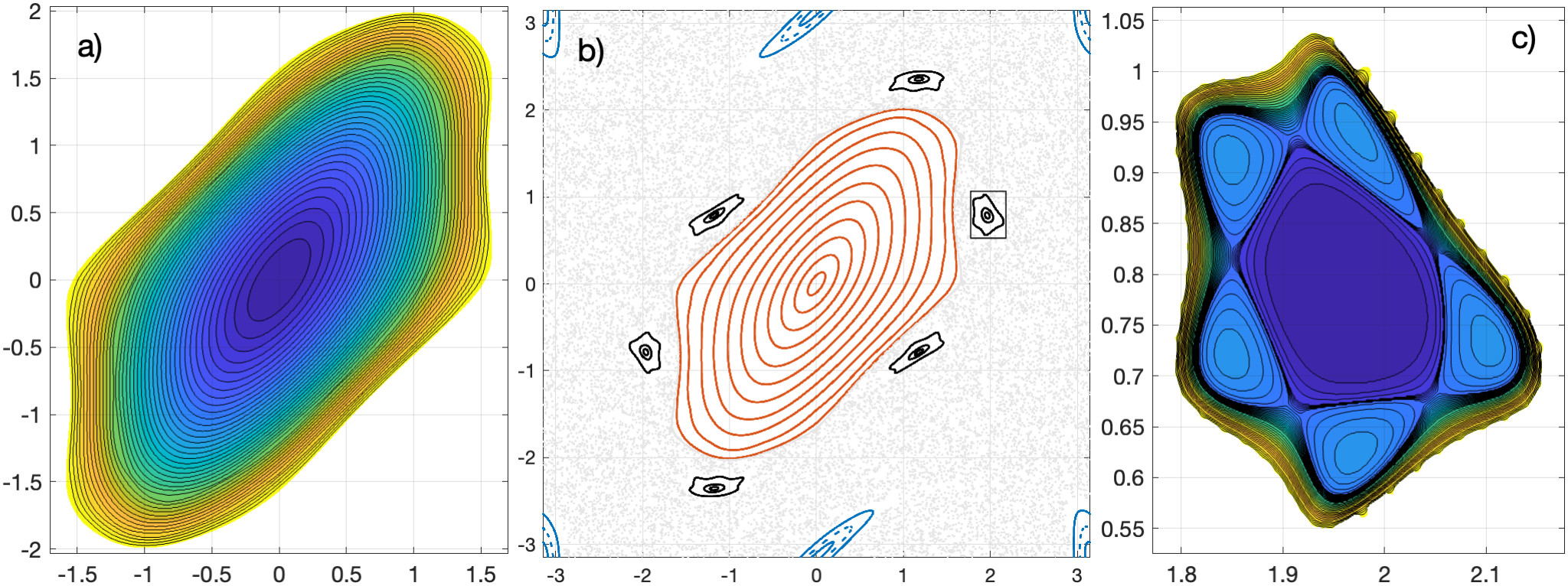}} 
\caption{
Local integrable structure of the standard map at $K=1.7$.
(a) Effective Hamiltonian $H_{\mathrm{eff}}(x,y)$ reconstructed on the period--1 island, shown as filled contours.
(b) Poincar\'e section of the map, with chaotic trajectories in gray and selected invariant curves outlining the period--1 (red), period--2 (blue) and period--6 (black) island chains.
(c) Effective Hamiltonian $H_{\mathrm{eff}}(x,y)$ reconstructed on a single lobe of the period--6 chain shown zoomed on the black box in panel (b).
In panels (a) and (c), $H_{\mathrm{eff}}$ is normalized to $[0,1]$ and displayed only on the corresponding regular region.
}
\label{fig:phase_portrait}
\end{figure*}

Temporal periodicity allows the dynamics to be expressed in terms of the one-period 
evolution operator $U$,
\begin{equation}
\rho(r, t + T) = U(T) \rho(r, t).
\end{equation}
Writing $\rho(r,t) = e^{\mu t}\phi(r,t)$ with $\phi$ periodic in time yields the 
eigenvalue problem
\begin{equation}
U \phi = e^\mu \phi \equiv \lambda \phi,
\label{eq:eigenvalue}
\end{equation}
so that the spectrum of $U$ determines decay rates and spatial structure of all modes. 
Because diffusion renders $U$ non-unitary, $|\lambda| < 1$. Throughout this work we 
characterize eigenvalues in terms of the decay rate
\begin{equation}
\gamma = -\frac{1}{T}\log|\lambda|,
\end{equation}
which reduces to $\gamma = -\log|\lambda|$ for the unit period $T=1$ used in all 
computations. The dominant eigenvalue governs the asymptotic decay, but at finite 
times the scalar field typically reflects contributions from multiple eigenmodes; 
this becomes increasingly important as $\mathrm{Pe}$ grows and spectral gaps shrink.

The stroboscopic dynamics admits a split-operator representation. Starting from 
Liouville's equation
\begin{equation}
\frac{\partial\rho}{\partial t} + i\hat{L}(t)\rho = 0,
\end{equation}
with Liouvillian
\begin{equation}
i\hat{L}(t) = \frac{\partial\psi}{\partial y}\frac{\partial}{\partial x} - 
\frac{\partial\psi}{\partial x}\frac{\partial}{\partial y} = \vec{u} \cdot \nabla,
\end{equation}
the delta-function forcing yields the one-period advective operator
\begin{equation}
\rho^{(t+T)}(x, y) = \exp\left(-y\frac{\partial}{\partial x} - K \sin x 
\frac{\partial}{\partial y}\right) \rho^{(t)}(x, y).
\end{equation}
Including diffusion and working in Fourier space gives the full one-cycle 
advection--diffusion operator in the form
\begin{equation}
\rho^{(t+T)}(m, n) = \sum_{k=-\infty}^{\infty} J_{m-k}(nK) \rho^{(t)}(k, k+n) 
e^{-D(m^2+n^2)},
\end{equation}
where $J_\ell$ denotes a Bessel function of the first kind. The diffusive factor 
exponentially suppresses large wavenumbers, permitting truncation to a finite 
Fourier basis. As $D$ decreases, the required truncation grows rapidly, making 
the large-$\mathrm{Pe}$ regime computationally demanding. 

Our objective is to compute and interpret a substantial portion of the spectrum of the
advection--diffusion operator at large Péclet numbers.
We compute leading eigenpairs using an implicitly restarted Arnoldi method, and show that
the resulting eigenmodes organize into distinct diffusive, advective, and mixed families
whose scaling, multiplicity, and ordering are dictated by the underlying Lagrangian
phase-space geometry. Details of the numerical implementation, including convergence 
diagnostics and resolution considerations, are given in Appendix A.

\section{Phase-Space Structure at $K=1.7$}

We focus on the standard map at $K=1.7$, a prototypical mixed phase space. At this 
parameter value, three prominent island chains of period $1$, $2$, and $6$ coexist 
with a substantial chaotic sea. Figure~\ref{fig:phase_portrait} shows this structure 
and introduces the geometric objects that organize the spectrum.

Panel~(b) shows a Poincaré section of the map, with chaotic trajectories in gray and 
selected invariant curves outlining the dominant regular regions. The period-1 island 
occupies the central region, while smaller chains of period 2 and period 6 are embedded 
in the surrounding chaotic sea. These regular regions form the non-chaotic components of 
the Lagrangian dynamics.

Panels~(a) and~(c) display effective Hamiltonians $H_{\mathrm{eff}}(x,y)$ reconstructed 
on representative regular regions: the period-1 island and a single lobe of the period-6 
chain, respectively. In each case, $H_{\mathrm{eff}}$ provides a local integrable 
surrogate for the map, with level sets approximating invariant curves in the corresponding 
region. The period-1 island appears as a smooth, nearly quadratic well, whereas the 
period-6 lobe exhibits pronounced non-quadratic structure associated with internal 
resonances. Details of the numerical reconstruction of the local Hamiltonian structure are given in Appendix B.

\section{Finite-Time Dynamics at Large Péclet Number}

To motivate the need for a spectral description beyond the dominant eigenmode, we examine 
the finite-time evolution of simple, localized disturbances under the advection-diffusion 
operator at $D=10^{-6}$. We focus on global homogenization: how a localized patch of 
scalar, initially in the chaotic sea, is transported and injected into regular components 
of phase space.

We consider two initial conditions constructed from the same compactly supported Gaussian 
$G(x,y)$ centered at $(x_0,y_0)$,
\[
\rho_1(x,y,0) = \partial_x G(x,y), \qquad
\rho_2(x,y,0) = \partial_y G(x,y).
\]
Both fields are localized on the same spatial scale, have zero spatial mean, and are 
supported in the same region of the chaotic sea. They differ only by a $\pi/2$ rotation in 
the orientation of their gradients. Any difference in their long-time behavior must 
therefore arise not from where the scalar is placed, but from how the early-time advective 
dynamics transports it into regular sets.

\begin{figure*}

\includegraphics[width=\textwidth]{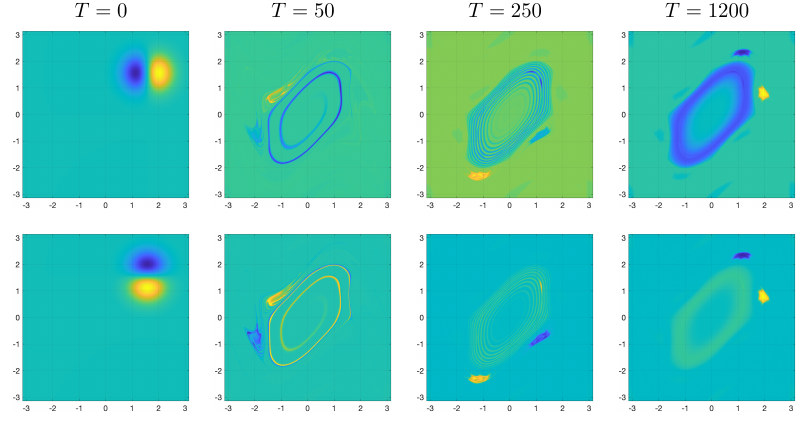}
\caption{Evolution of two localized initial conditions under the advection-diffusion 
operator for $K = 1.7$, $D=10^{-6}$. The two rows correspond to 
$\rho_1=\partial_x G$ (top) and $\rho_2=\partial_y G$ (bottom), where $G$ is the same 
Gaussian centered at $(x_0,y_0)$. Columns show the initial condition and the fields after 
$n=50$, $250$, and $1200$ periods.}
\label{Fig3}
\end{figure*}

Figure~\ref{Fig3} shows the evolution of these two initial conditions over $1200$ periods. 
In both cases, the scalar is rapidly homogenized within the chaotic sea and progressively 
accumulates inside the primary elliptic island. At late times the fields occupy the same 
geometric support: the same island core, the same surrounding invariant manifolds, and the 
same hierarchy of filaments. What differs is not which structures are present, but the 
relative amplitudes with which they are populated. These weights are set during the early 
advective stage, when material originating in the chaotic region is injected into regular 
regions in a manner that depends sensitively on the orientation of the initial gradients.

\begin{figure}

\includegraphics[width=0.4\textwidth]{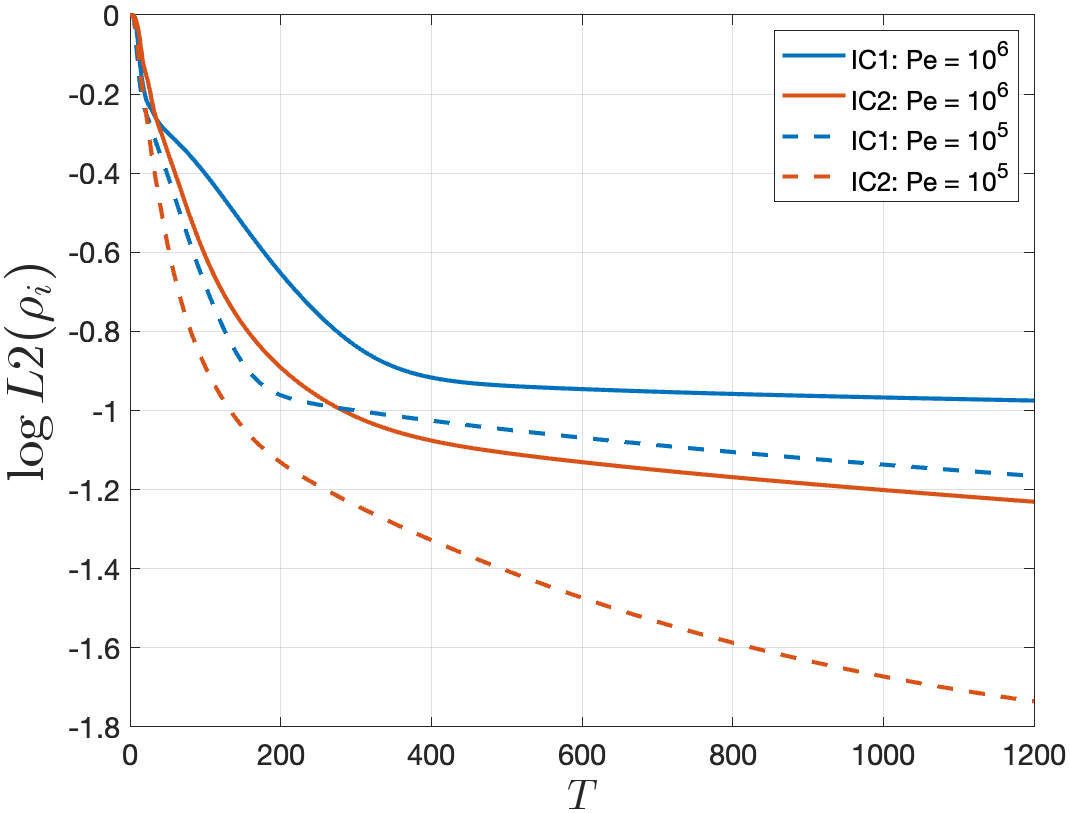}
\caption{Time evolution of the $L^2$ norm of the scalar fields shown in 
Fig.~\ref{Fig3}. Solid curves correspond to $\mathrm{Pe}=10^6$, dashed curves to 
$\mathrm{Pe}=10^5$. Blue curves show $\rho_1=\partial_x G$; red curves show 
$\rho_2=\partial_y G$.}
\label{Fig3a}
\end{figure}

This distinction is dynamically relevant. Figure~\ref{Fig3a} shows that the $L^2$ norms of 
the two solutions decay at persistently different rates over the entire accessible time 
window. Even after $10^3$ periods---orders of magnitude longer than any advective time 
scale---the system has not collapsed onto a unique decay rate. Although both solutions 
occupy the same geometric subspace, they project onto different combinations of long-lived 
modes.

The implication is fundamental: at large Péclet number and finite times, the dynamics is 
not governed by a single dominant eigenmode. What remains relevant is a collection of 
slowly decaying modes associated with the same regular structures, whose relative weights 
are determined by how the initial density is transported from the chaotic sea into those 
regions. Understanding the dynamics therefore requires access to a leading portion of the 
spectrum and to the organization of these long-lived families.

\section{Global Structure of the Computed Spectrum}

For each Péclet number we compute a fixed number of leading eigenpairs 
$U\phi=\lambda\phi$, where $U$ advances the scalar field by one forcing period. 
Because $U$ is non-unitary for $D>0$, all eigenvalues satisfy $|\lambda|<1$. 
We represent the spectrum in polar coordinates
\begin{equation}
\theta = \arg(\lambda), \qquad \gamma = -\log|\lambda|,
\end{equation}
where $\gamma$ is the exponential decay rate and the unit circle maps to $\gamma=0$. 
This transformation unwraps the complex plane into a strip in which eigenvalues at 
different Péclet numbers can be compared on a common scale.

\begin{figure*}

\centering
\includegraphics[width=0.85\textwidth]{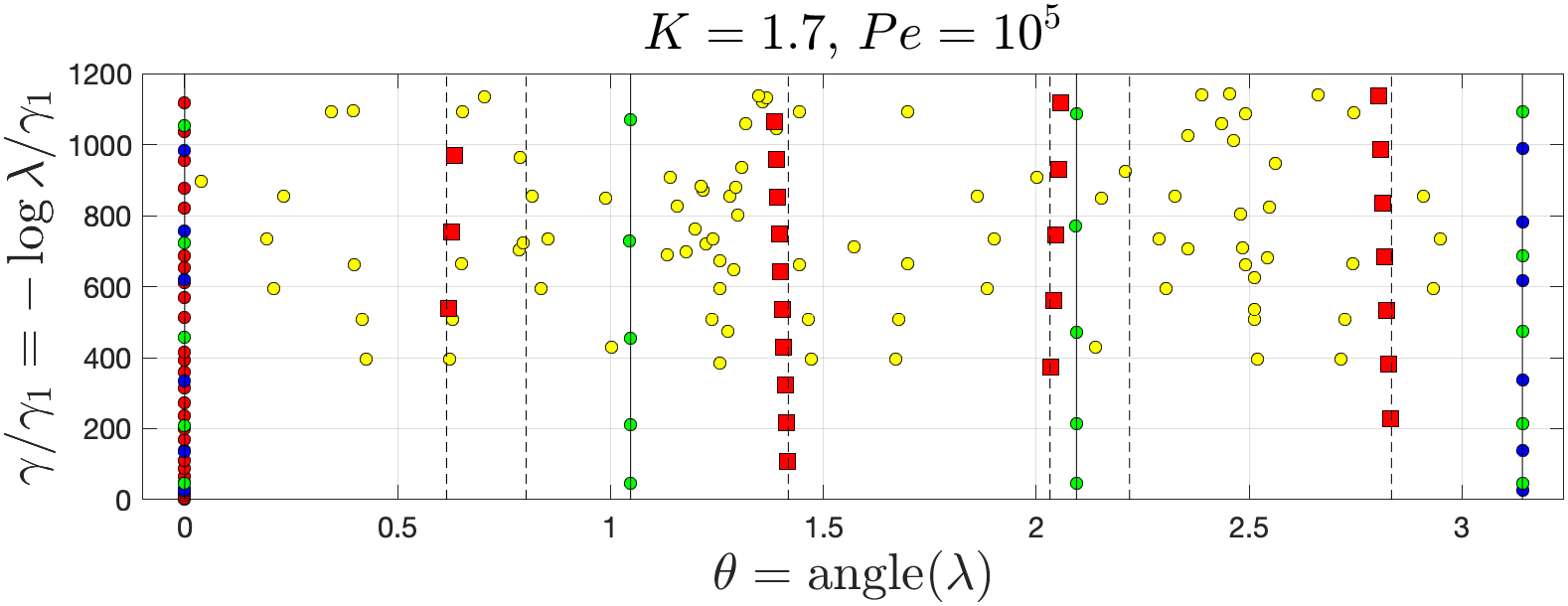}
\includegraphics[width=0.85\textwidth]{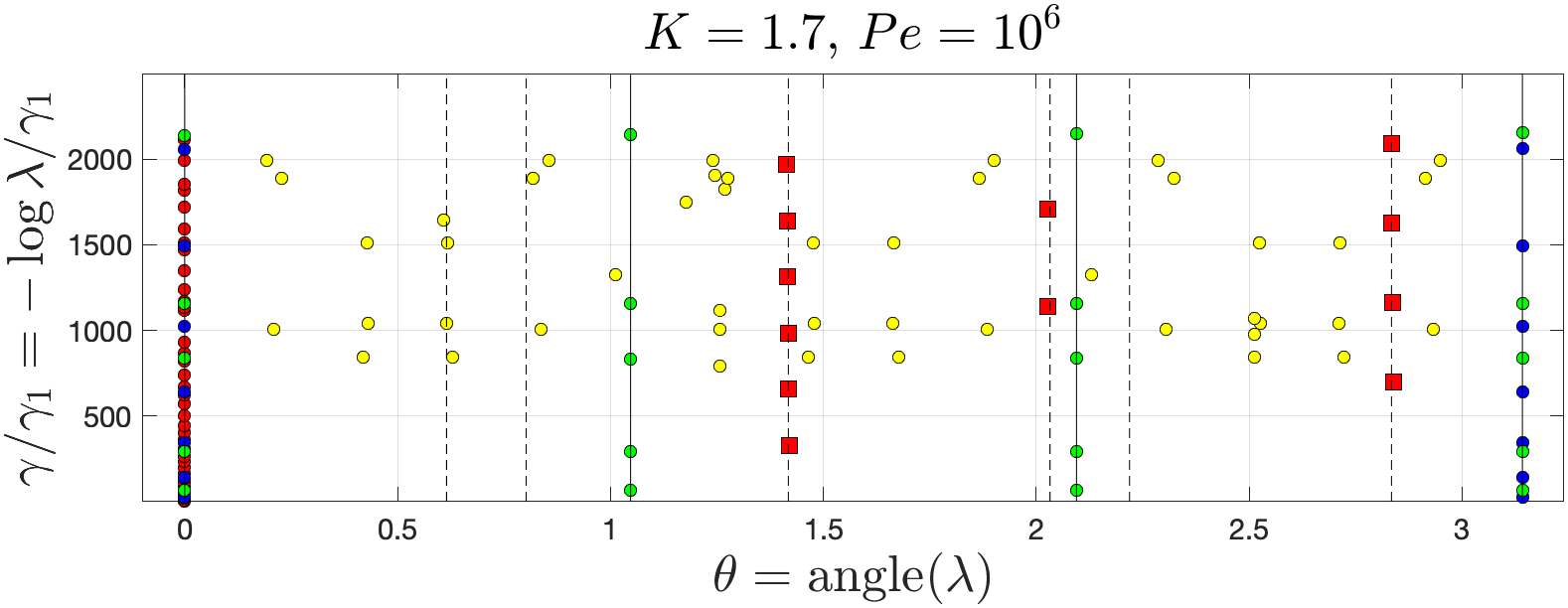}
\includegraphics[width=0.85\textwidth]{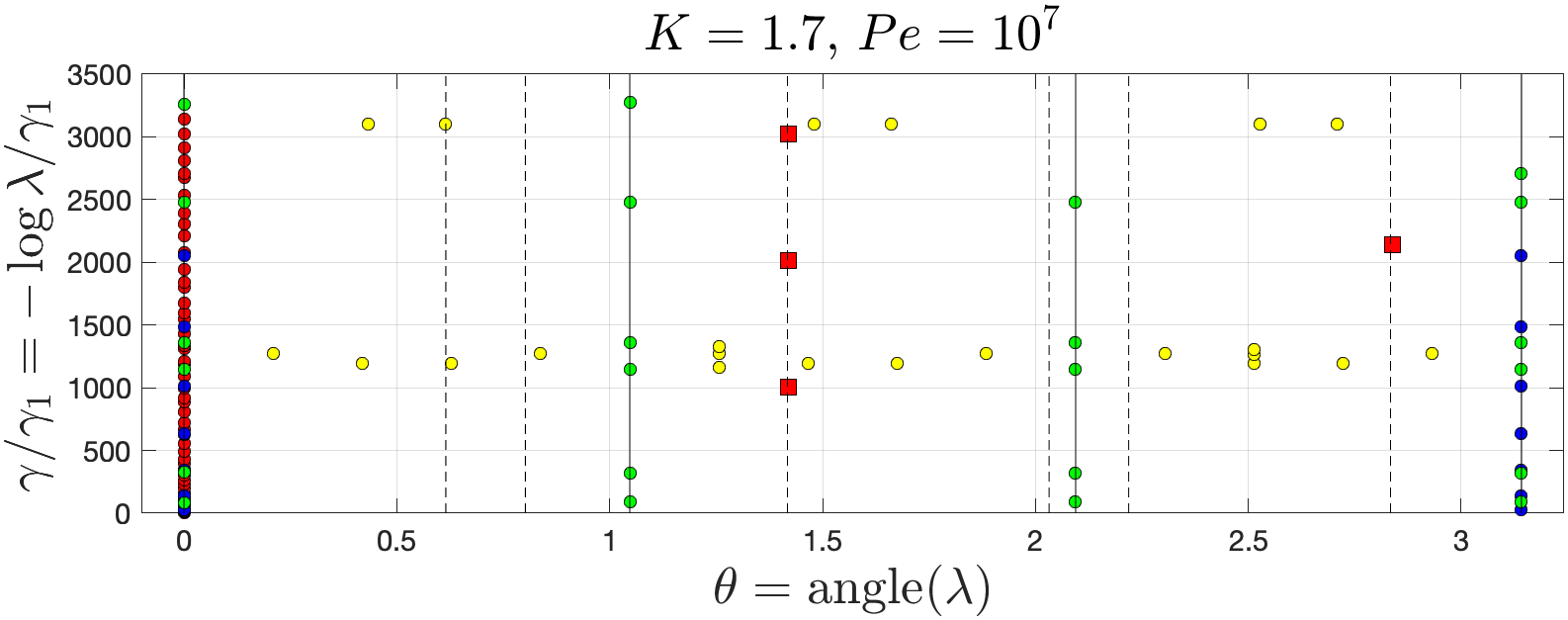}
\caption{
Leading eigenvalues of the one-period advection-diffusion operator for 
$\mathrm{Pe}=10^5,\,10^6,$ and $10^7$, plotted in polar form using 
$\theta=\arg(\lambda)\in[0,\pi]$ (folded by complex-conjugate symmetry) and 
$\gamma=-\log|\lambda|$, normalized in each panel by the leading decay rate $\gamma_1$. 
Solid vertical lines mark the symmetry directions $\theta = m\pi/3$ ($m=0,1,2,3$), 
corresponding to the roots of unity associated with the dominant period-1, 2, and 6 island 
chains. Colored circles denote diffusive eigenmodes $\psi^{p}_{m,k}$, grouped by island 
family $p$ ($p=1,2,6$) and clustered along these symmetry directions, forming discrete 
diffusive ladders. Red squares identify advective modes $\phi$, which lie away from the 
symmetry axes and populate the spectral gaps between diffusive branches. Dashed vertical 
lines indicate the independently computed advective phases obtained from the mean rotation 
number $\rho(I)$ of invariant tori within the period-1 island.}
\label{spectra}
\end{figure*}

Figure~\ref{spectra} shows the leading spectra for $\mathrm{Pe}=10^5,\,10^6,$ and 
$10^7$ in these coordinates. Three features emerge immediately.
First, the spectrum is far from a featureless cloud. Eigenvalues organize strongly in 
phase, with pronounced vertical alignments that persist across all three Péclet numbers. 
The solid vertical guide lines mark $\theta=0,\pi/3,2\pi/3,$ and $\pi$, corresponding to 
the dominant period-1, 2, and 6 island chains visible in the phase portrait. These angles 
are not arbitrary plotting artifacts but are tied to geometric structures in the underlying 
map.

Second, the real axis $\theta=0$ plays a special role. Eigenvalues associated with all 
three island periods appear there, so this column does not correspond to a single 
dynamical mechanism or period. Distinct geometric origins are already interleaved in 
decay rate at the most prominent phase.

Third, increasing Péclet number does not simply sharpen a fixed set of branches. As $D$ 
decreases, additional eigenvalues appear increasingly close to the unit circle, populating 
phase angles that were previously sparse or empty. Because we compute a fixed number of 
leading eigenpairs at each $\mathrm{Pe}$, these plots represent a moving window into an 
ever denser near-unit-circle spectrum: higher Péclet numbers reveal more angular variety, 
while the apparent prominence of any individual alignment reflects sampling rather than 
intrinsic importance.

As $D\to0$, the advection operator admits continuous spectral components on the unit 
circle, and the discrete spectrum of the diffusive problem progressively resolves this 
dense set of phases. At finite $D$, however, the spectrum remains discrete, and the 
question is how these accumulating eigenvalues organize in relation to the underlying 
phase-space geometry.

\section{Spectral Families and Universality in Mixed Phase Space}

The spectra in Fig.~\ref{spectra} reveal distinct, persistent branches. What appears at 
first glance as a dense cloud of eigenvalues separates into organized families. Modes 
clustered along the symmetry directions $\theta=m\pi/3$ form discrete ladders, 
color-coded in Fig.~\ref{spectra} by island family and identified in the caption as 
diffusive modes $\psi^{p}_{m,k}$. Between these axes lie advective modes $\phi$, 
occupying distinct phases from those of the diffusive branches.

Eigenvalues that align along a given branch correspond to eigenfunctions with closely 
related spatial structure and common dynamical origin, while modes in different regions of 
the spectrum exhibit qualitatively different morphology. The spectrum therefore decomposes 
naturally into a small number of spectral families, each tied to a distinct transport 
mechanism in the mixed phase space. We make this classification explicit and show that the 
internal organization of each family is universal across $\mathrm{Pe}$.

\begin{figure*}

\includegraphics[width=\textwidth]{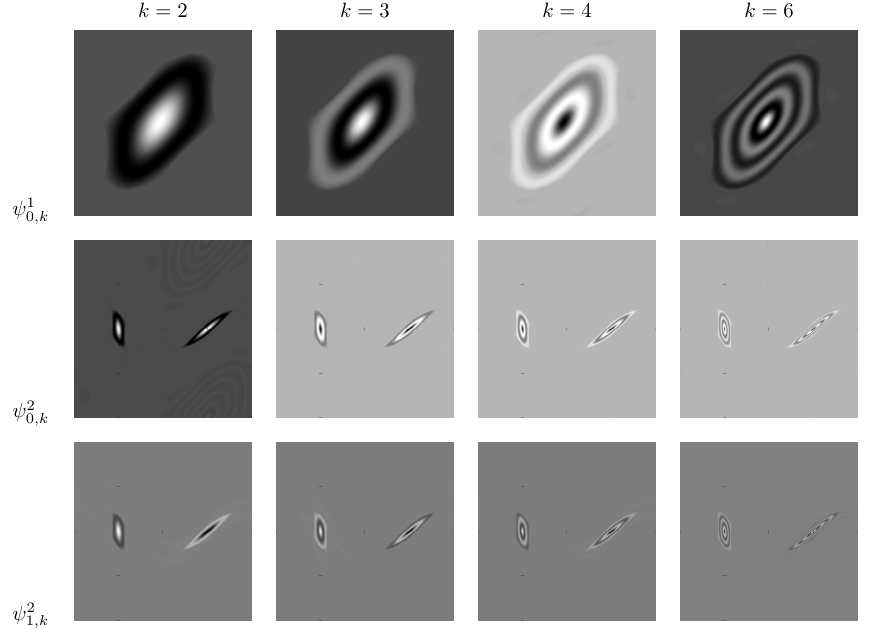}
\caption{
  Three families of diffusive modes for $K = 1.7$, $D = 10^{-6}$. 
  Top row: Period-1 modes $\psi^1_{0,k}$ at $k = 2, 3, 4, 6$.
  Bottom two rows: Period-2 modes $\psi^2_{m,k}$ at the same $k$-values 
  in both symmetry classes ($m = 0, 1$) (For clarity, Period-2 modes are unwrapped on the torus and plotted in ($\tilde{x} = x +\pi/2, \tilde{y} = y -\pi)$).
}
\label{dmode1}
\end{figure*}

\subsection{Diffusive Modes}

We begin with the modes clustered along the symmetry axes in Fig.~\ref{spectra}, which are 
governed primarily by diffusion within regular islands.

\subsubsection{Island-localized modes}

\begin{figure*}

\includegraphics[width=\textwidth]{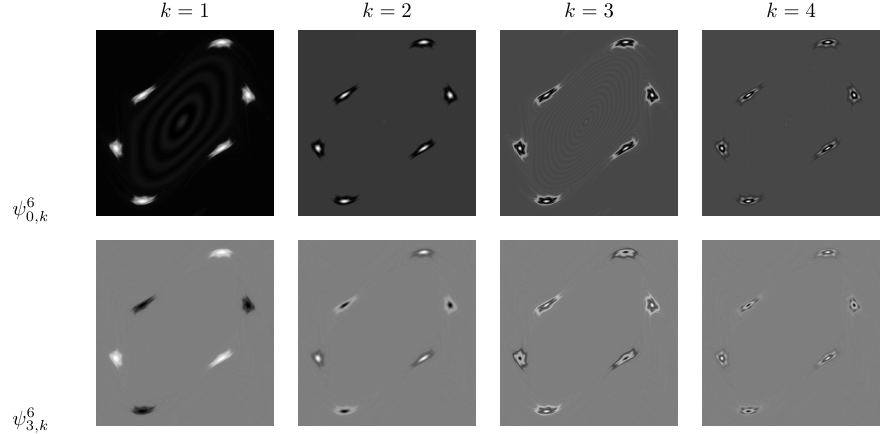}
\caption{
  Leading diffusive modes in the period 6 island chain for $K = 1.7$, $D = 10^{-6}$ for 
  $m=0,3$ symmetry classes.
}
\label{dmode2}
\end{figure*}

Figures~\ref{dmode1} and~\ref{dmode2} show representative eigenfunctions supported on 
individual regular islands. These modes occupy the full extent of a single island and 
closely resemble eigenfunctions of the Laplacian on a bounded domain. Their nodal patterns 
form approximately concentric bands across the island interior, with successive modes 
distinguished primarily by the number of such bands. In this sense, each regular island 
acts as a finite square well for the diffusive dynamics.

Dirichlet boundary conditions emerge naturally from the dynamics. Trajectories in the 
chaotic sea decorrelate rapidly and homogenize $\rho$ on timescales short compared with 
diffusion across an island. From the perspective of the island interior, the exterior 
therefore acts as a nearly uniform reservoir. Diffusive structure within the island is 
continually leaked into this well-mixed background, so that the island boundary behaves, 
to leading order, like a fixed-value interface. Each regular island thus forms a deep 
diffusive well embedded in a homogenizing environment, and its island-filling modes are 
naturally organized as eigenfunctions of a Dirichlet Laplacian on a bounded domain.

A period-1 island, a period-2 island, and a period-6 island all support this same 
square-well family, even though the spatial extent and layout of the islands differ. 
Within any given island, the dominant organizing parameter is therefore a radial mode 
index.

For islands of period $p>1$, this family further splits into $p$ symmetry classes, 
reflecting how the mode is phased across the $p$ dynamically equivalent copies of the 
island. A period-2 island, for example, admits two such classes, corresponding to global 
phases $0$ and $\pi$ under interchange of the two lobes. More generally, 
$m\in\{0,\dots,p-1\}$ labels these discrete phase classes.

We denote this family by  $\psi^{p}_{m,k}$
where $p$ is the island period, $k$ counts radial nodes, and $m$ labels the discrete phase 
class associated with the $p$-fold geometry.

Diffusion across a bounded region imposes Dirichlet scaling,
\begin{equation}
\gamma_k(\psi^p) \;\sim\; D\,k^2 .
\label{eq:psi_scaling}
\end{equation}

\begin{figure}

\includegraphics[width=0.5\textwidth]{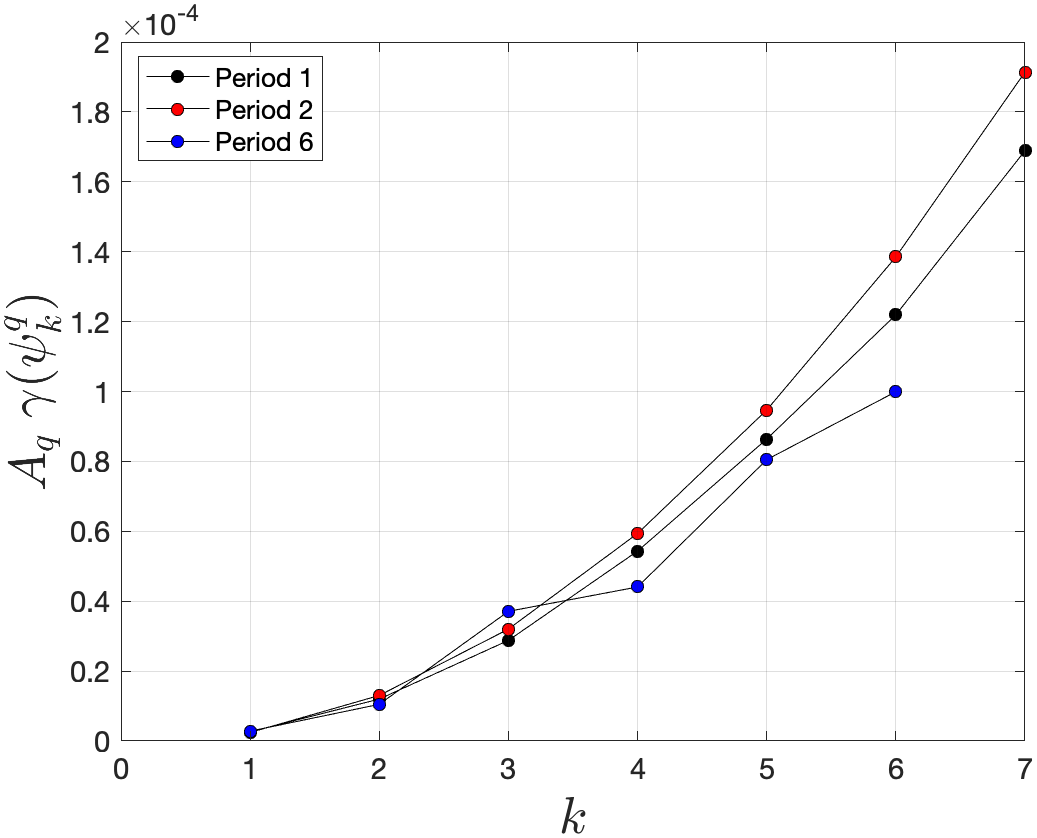}
\caption{Area-rescaled diffusive ladders at $\mathrm{Pe}=10^7$ for the period-1, 2, and 6 
island chains.}
\label{scaleddiff}
\end{figure}

Figure~\ref{scaleddiff} shows the isolated diffusive ladders for the dominant period-1, 2, 
and 6 island chains at $\mathrm{Pe}=10^7$ where the decay rates are multiplied by the 
computed area of single lobe each island chain. As expected for Laplacian-like modes on 
bounded domains, each ladder is approximately quadratic in the radial index $k$, with the 
overall scale set by island size,
\[
\gamma_k^{(p)} \;\sim\; D\,\frac{k^2}{A_p},
\]
up to a shape factor. Smaller islands therefore exhibit faster decay at fixed $k$ and 
steeper ladders, consistent with the ordering $A_1 \gg A_2 \gg A_6$. The shorter extent of 
the period-2 and period-6 ladders reflects weaker confinement: fewer well-localized 
diffusive levels remain before leakage into the chaotic sea and inter-island interactions 
become significant.

Equation~\eqref{eq:psi_scaling} is an asymptotic statement. The single radial index $k$ is 
only a coarse proxy for the two-dimensional geometry of a finite island, and the effective 
Laplacian felt by these modes depends on details of the island boundary that vary weakly 
with $D$. At small $k$ in particular, curvature, anisotropy, and boundary structure 
introduce order-one corrections to the simple square-well picture. The ladder 
$\{\psi^p_{m,k}\}$ should therefore be understood as an organizing framework, not as an 
exact spectrum of an isolated operator.

More importantly, the phase space contains multiple regular island chains, each supporting 
its own diffusive ladder. Nothing in the construction of these families enforces a 
spectral separation between ladders associated with different islands. It is therefore 
inevitable that, for some pairs of indices,
\[
\gamma(\psi^p_{m,k_p}) \;\approx\; \gamma(\psi^q_{m,k_q}),
\]
even when the underlying islands are dynamically distinct. These near coincidences in 
decay rate set the stage for interactions between ladders that lie beyond the 
single-island picture.

\subsubsection{Inter-island hybridization and avoided crossings}

The diffusive ladders $\{\psi^p_{m,k}\}$ described above are defined by localization on a 
single regular island chain and by the quadratic scaling 
$\gamma \sim \alpha_{p,m} D k^2$. This description is asymptotically correct in the limit 
of weak coupling between dynamically distinct regions. At finite $\mathrm{Pe}$, however, 
the spectrum is not a direct sum of independent island ladders. As $k$ increases, the 
decay rates of different ladders inevitably approach one another, and the assumption of 
isolation breaks down.

When two diffusive levels,
\[
\gamma(\psi^1_{0,k}) \;\approx\; \gamma(\psi^p_{m,k_p}),
\]
come into near degeneracy, the corresponding eigenfunctions no longer remain confined to 
their parent islands. Instead, the eigenvalue problem produces a pair of modes whose 
support spans both regular regions. Each member of the pair carries comparable weight on 
the two islands, but with opposite relative phase. In direct analogy with symmetric and 
antisymmetric combinations in tunneling problems, the near-degeneracy is lifted into two 
distinct eigenmodes with closely spaced decay rates.

\begin{figure}

\includegraphics[width=0.5\textwidth]{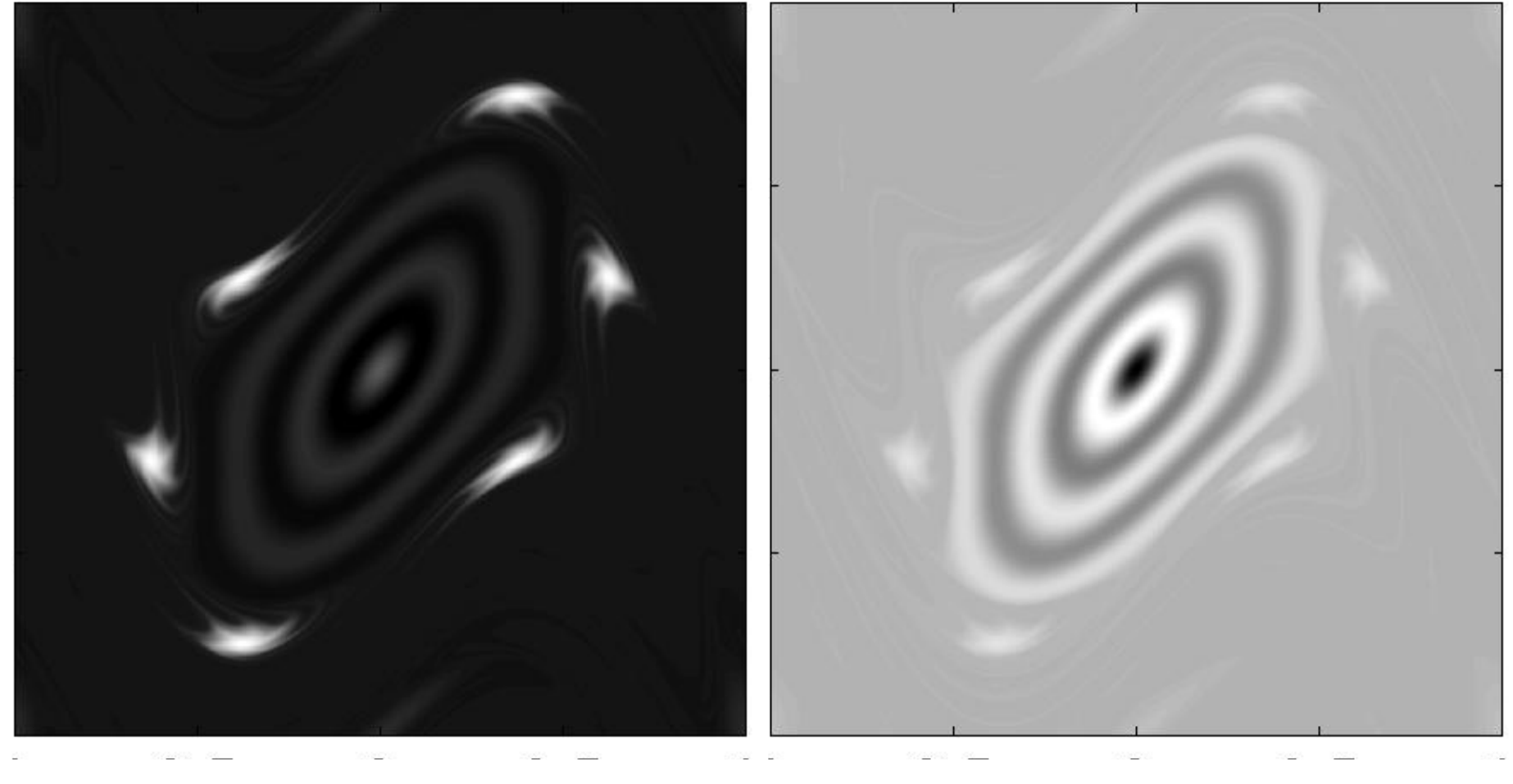} \\
\includegraphics[width=0.5\textwidth]{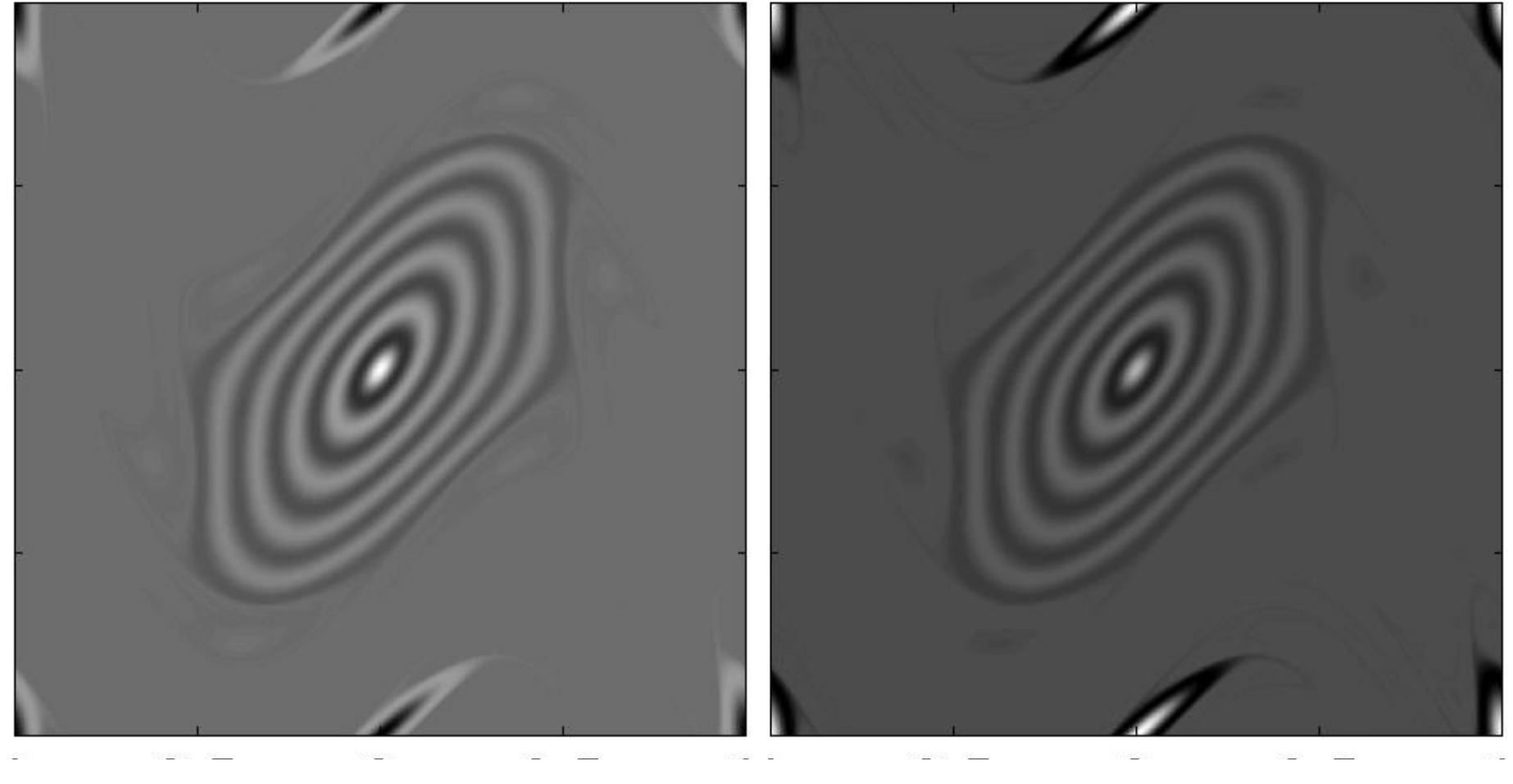} \\
\includegraphics[width=0.5\textwidth]{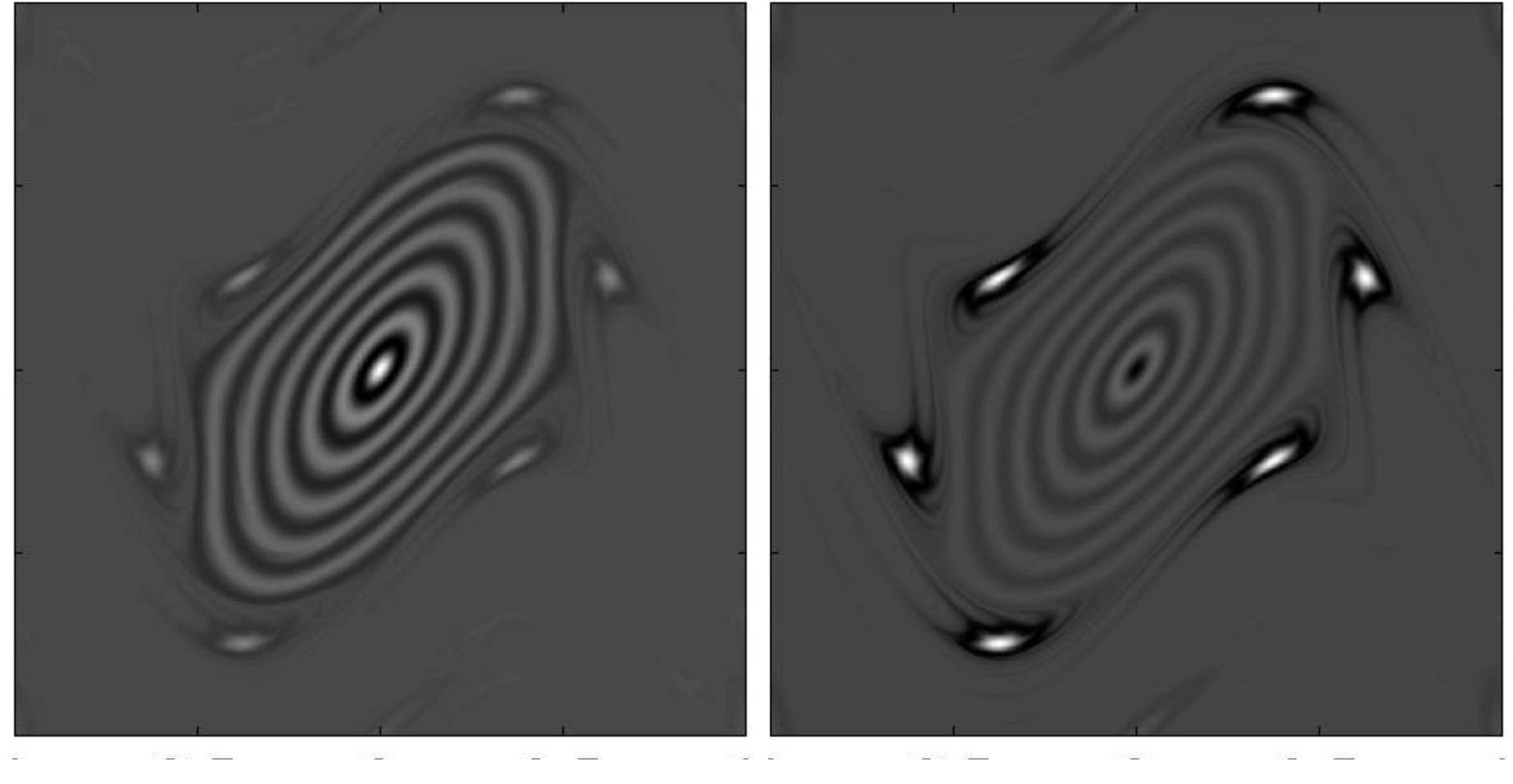}
\caption{Leading pairs of hybrid (tunneling) diffusive eigenmodes spanning multiple 
islands at $\mathrm{Pe} = 10^5$. Each row shows the two hybrid eigenmodes associated with 
a single period-1 diffusive mode, $\psi^1_{0,k}$, split by coupling to a diffusive mode 
in the secondary island chain. Top row: $\psi^1_{0,6}/\psi^6_{0,1}$ pair. Middle: 
$\psi^1_{0,10}/\psi^2_{0,2}$. Bottom: $\psi^1_{0,12}/\psi^6_{0,2}$ pair.}
\label{mixed_modes}
\end{figure}

Figure~\ref{mixed_modes} shows the leading examples of this process at 
$\mathrm{Pe}=10^5$. Each row corresponds to a single period-1 diffusive ladder level 
$\psi^1_{0,k}$ that has entered near resonance with a diffusive level on a secondary 
island chain. The two panels in a row are the split descendants of that one ladder level: 
hybrid eigenmodes that differ only in their relative phase across islands. What would be a 
single diffusive mode in the uncoupled limit thus appears in the full problem as a paired 
structure.

These hybrid modes mark the breakdown of the single-island ladder picture. They are 
neither members of the period-1 family nor of the secondary ladder alone, and they do not 
obey either island's diffusive scaling law. Spectrally, they occupy the narrow gaps opened 
by avoided crossings between ladders. Geometrically, they represent tunneling of diffusive 
structure across invariant barriers separating regular regions.

As $\mathrm{Pe}$ increases, the diffusive resolution scale shrinks and most 
near-degeneracies between diffusive ladders produce only weak leakage between branches. 
Hybridization is confined to a sparse, geometrically selected set of ladder levels. In the 
small-$D$ limit the spectrum is well approximated by a direct sum of island families. The 
only systematic exceptions occur at isolated $(k_1,k_p)$ pairs where geometric 
commensurability is most precise. At those locations, genuinely mixed eigenmodes persist 
even as $D\!\to\!0$.

\begin{figure}
\centering
\includegraphics[width=0.5\textwidth]{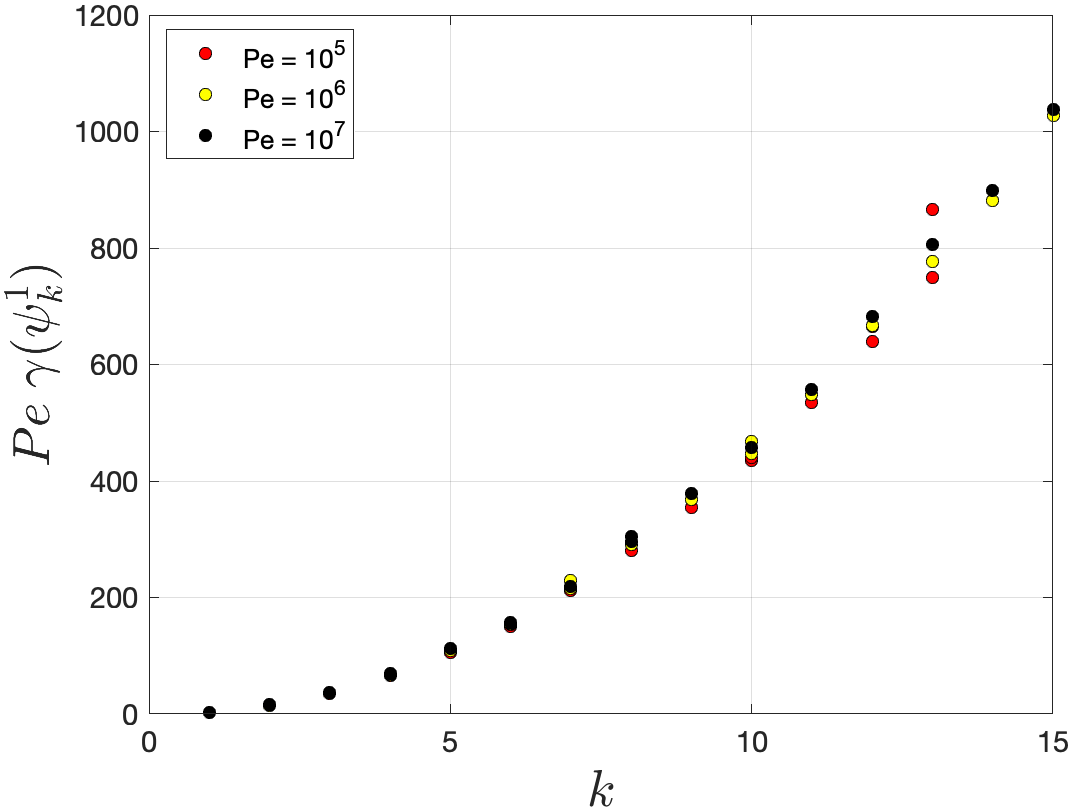}
\includegraphics[width=0.5\textwidth]{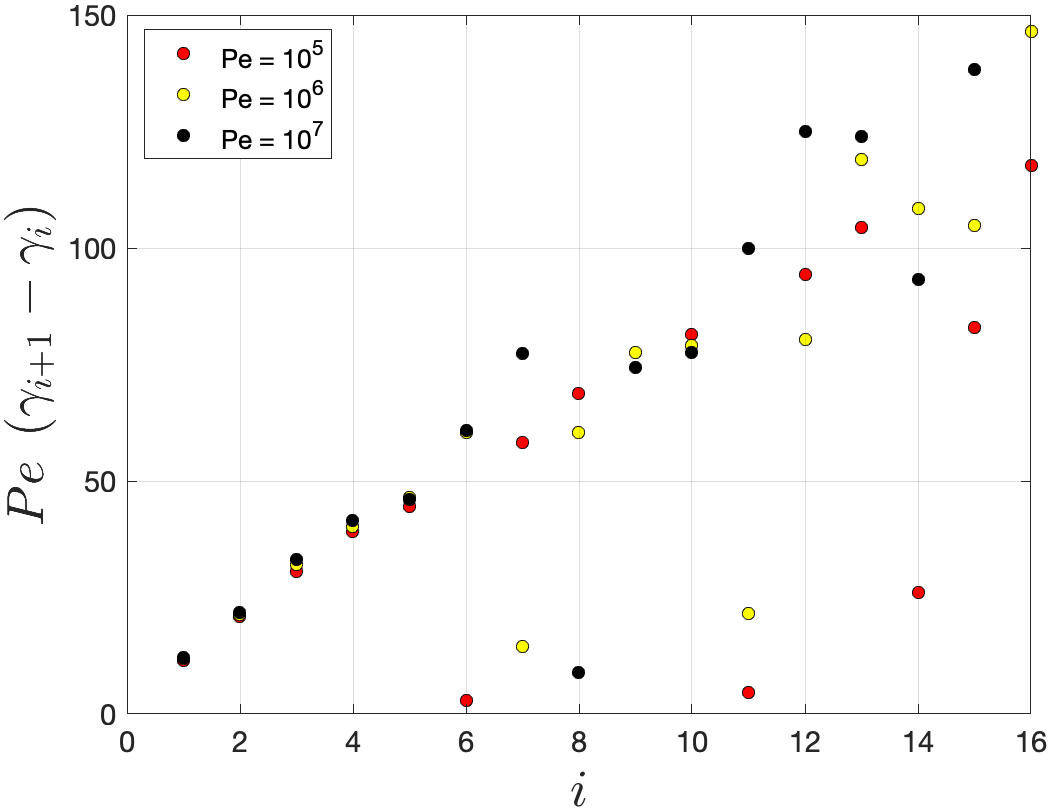}
\caption{
(top) Scaled decay rates $\mathrm{Pe}\,\gamma(\psi^1_k)$ for the period-1 diffusive 
family (including hybridized modes) at $\mathrm{Pe}=10^5,10^6,10^7$, plotted against the 
radial ladder index $k$. (bottom) Consecutive scaled gaps 
$\mathrm{Pe}\,(\gamma_{i+1}-\gamma_i)$ along the same branch, where the modes are ordered 
by increasing $\gamma$.
}
\label{fig:dspectra}
\end{figure}

\subsubsection{Diffusive Scaling}

Figure~\ref{fig:dspectra} shows two complementary aspects of the dominant diffusive 
branch. The upper panel shows that, once scaled by $\mathrm{Pe}$, the period-1 ladder 
collapses onto a single curve across three decades in $\mathrm{Pe}$. This collapse holds 
even in the presence of hybridized levels, confirming that the $\theta=0$ branch 
constitutes a single, universal diffusive family, consistent with the island localization 
demonstrated by Gorodetskyi et al.~\cite{Gorodetskyi:2012PoFa,Gorodetskyi:2012PoFb}. 
Hybridization modifies individual levels, but it does not destroy the identity of the 
ladder as a whole.

The lower panel reveals a more delicate consequence of inter-island coupling. Although the 
backbone of the ladder follows the expected diffusive trend, the spacing between 
successive modes is punctured by isolated, persistent collapses. These anomalously small 
gaps occur at fixed locations in the ordering for all $\mathrm{Pe}$ and correspond to 
hybridized levels. Thus, even on the simplest and most dominant branch of the spectrum, 
the presence of multiple island chains removes any control over spectral gaps. The law 
$\gamma\sim Dk^2$ governs the coarse organization of the ladder, but it does not determine 
its fine structure.

\subsection{Core-localized (advective) modes}

\begin{figure*}
\includegraphics[width=\textwidth]{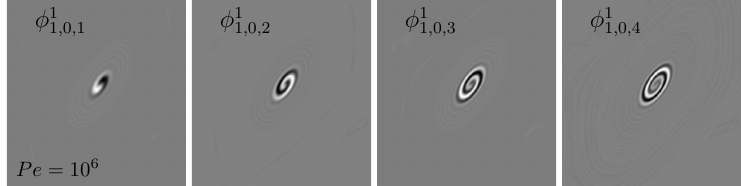} \\
\includegraphics[width=\textwidth]{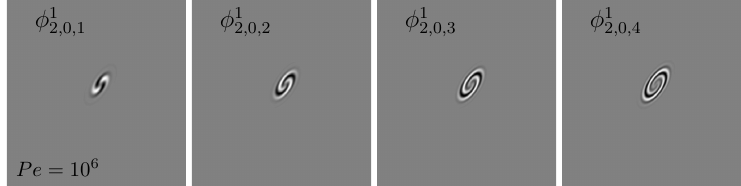} \\
\includegraphics[width=\textwidth]{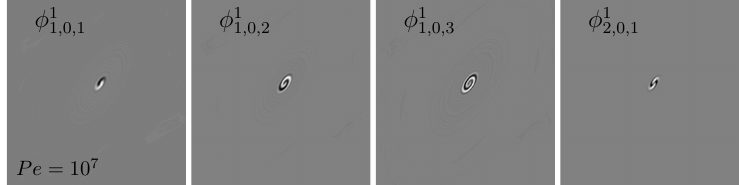}
\caption{
  Advective (core-localized) modes in the period-1 island for $K = 1.7$. 
  Top two rows: $D = 10^{-6}$, azimuthal families $\phi^1_{1,0,k}$ ($q=1$) 
  and $\phi^1_{2,0,k}$ ($q=2$) for $k = 1, 2, 3, 4$. 
  Bottom row: $D = 10^{-7}$, showing $\phi^1_{1,0,k}$ for $k = 1, 2, 3$ 
  and $\phi^1_{2,0,1}$.
}
\label{amodes}
\end{figure*}

A second, qualitatively distinct family is shown in Fig.~\ref{amodes}. These 
eigenfunctions are sharply localized near the elliptic cores of regular islands and form 
compact, approximately circular patterns centered on the fixed point. Their structure and 
spectral organization are those of a two-dimensional harmonic oscillator: the modes appear 
in short ladders, typically in complex-conjugate pairs, with successive members exhibiting 
additional radial rings.

In this sense, these eigenfunctions are local Koopman modes of the integrable dynamics in 
the island core~\cite{Mezic2013}. They represent a coherent rotation about the elliptic point, 
weakly damped by diffusion.

Like the island-filling modes, these eigenfunctions inherit a discrete phase label 
$m\in\{0,\dots,p-1\}$ associated with the $p$-fold island geometry. In addition, two 
internal indices are required to distinguish their oscillator structure: an azimuthal 
index $q$, which fixes the spectral phase of the eigenvalue, and a radial index $k$. We 
therefore denote this family by
\begin{equation}
\phi^{p}_{q,m,k}.
\end{equation}

The physical meaning of the additional index $q$ is rooted in the local Hamiltonian 
structure of an elliptic island core. In a neighborhood of an elliptic fixed point, the 
standard map is smoothly conjugate to a rotation in action-angle coordinates,
\[
(I,\vartheta) \;\mapsto\; (I,\vartheta + \Omega(I)),
\]
with $\Omega(I)$ approximately constant near the origin. In the absence of diffusion, 
scalar density in this region is simply transported by rigid rotation. The corresponding 
Koopman operator therefore admits Fourier modes $e^{iq\vartheta}$ as exact eigenfunctions, 
with eigenvalues $e^{iq\Omega}$.

Finite diffusivity introduces weak transverse leakage across level sets of $I$, converting 
these neutral Koopman modes into slowly decaying eigenfunctions of the advection-diffusion 
operator. The integer $q$ thus labels the azimuthal sector of the local rotation, while 
$k$ counts radial excitations within the resulting harmonic well. The spectral phase
\[
\theta \;=\; \arg(\lambda)
\]
is therefore not arbitrary: it encodes the underlying Hamiltonian rotation frequency of 
the island core. In the raw spectra, this structure appears as discrete vertical bands in 
$\theta$, one for each admissible value of $q$.

For weakly damped coherent rotation, diffusion acts only through slow transverse leakage, 
imposing linear scaling in the oscillator quantum number,
\begin{equation}
\gamma_k(\phi^p) \;\sim\; \sqrt{D}\,k .
\label{eq:phi_scaling}
\end{equation}
The prefactor carries a weak dependence on the azimuthal index $q$ through the local 
curvature of the island core: modes with larger $q$ experience stronger azimuthal shear 
and therefore enhanced leakage, giving the refined estimate
\[
\gamma_k \sim \sqrt{D}\,\sqrt{q}\,k .
\]
This geometric $q$-dependence will determine where advective ladders first intersect the 
diffusive spectrum in Sec.~VII.

\begin{figure}
\includegraphics[width=0.475\textwidth]{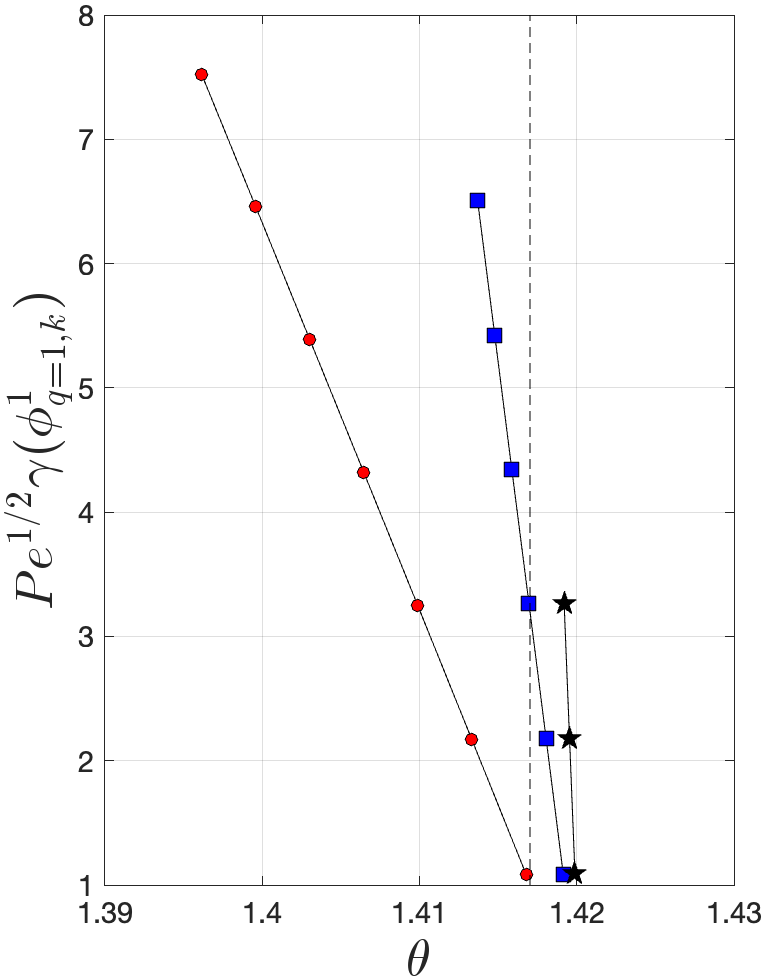}
\includegraphics[width=0.475\textwidth]{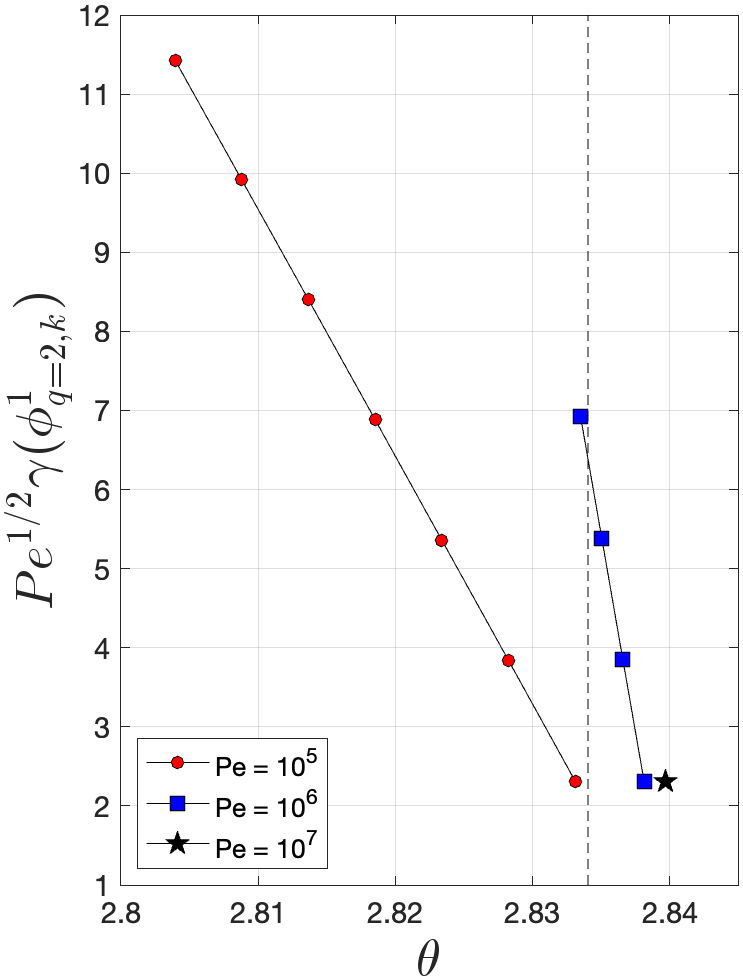}
\caption{Rescaled decay rate $\mathrm{Pe}^{1/2}\,\gamma$ versus spectral phase $\theta$ 
for the dominant advective ladders in the period-1 island. Left: $q=1$ branch. Right: 
$q=2$ branch. The dashed line shows the geometric prediction $q\,\theta_{\mathrm{adv}}$ 
obtained from the local island rotation frequency.}
\label{aspect}
\end{figure}

Figure~\ref{aspect} isolates the dominant advective branches from the raw spectrum, 
rescales the decay rate by $\mathrm{Pe}^{1/2}$, and displays the result in a narrow 
window of spectral phase $\theta$ for the primary azimuthal sectors $q=1$ (left) and 
$q=2$ (right). In these coordinates each family collapses onto a single slanted ladder, 
with successive $k$ forming the expected harmonic-oscillator hierarchy.

Across three decades in Péclet number the points align rigidly and the spacing in $k$ is 
preserved, confirming the $\mathrm{Pe}^{-1/2}$ scaling predicted for advective 
(core-localized) modes~\cite{Vukadinovic2015,bedrossian2017}. The dashed line in each panel marks the geometric phase 
$q\,\theta_{\mathrm{adv}}$ obtained directly from the local island rotation frequency; the 
observed branches follow this prediction without any spectral fitting. The small drift of 
$\theta$ toward smaller values with increasing $k$ reflects azimuthal shear in the island 
core: higher $k$ modes extend farther in radius and therefore experience slightly 
different rotation rates. Because diffusion spreads the modes across streamlines, the 
shear-induced phase shift decreases with increasing $\mathrm{Pe}$ as this cross-stream 
spreading weakens.

\section{Geometric Prediction of Spectral Ordering}

The classification developed above does more than organize the observed spectrum: it
provides a concrete framework for predicting where distinct families of modes must
appear in the ordered eigenvalue list.  The slow spectrum is assembled from a finite
collection of square--well ladders $\psi^{p}_{k,m}$, one for each regular island chain,
together with harmonic--oscillator ladders $\phi^{p}_{q,m,k}$ associated with elliptic
cores.  Each ladder obeys a simple asymptotic law,
\[
\gamma^{(p)}_k \;\sim\; D\,\alpha_p k^2, \qquad
\gamma^{(\phi)}_k \;\sim\; \sqrt{D}\,\beta_p k,
\]
with prefactors $\alpha_p$ and $\beta_p$ determined by geometric properties of the
corresponding island: effective width for the square--well modes and local curvature of
the Hamiltonian near the elliptic point for the oscillator modes.

\medskip

\subsection{Geometric predictors for inter--island hybridization}
\label{sec:prediction}

Within each regular island the diffusive spectrum organizes into a ladder of modes.
The predictor developed here is based entirely on geometry and scaling, not on the
spectrum itself.  In a period--$p$ island of area $A_p$, the $k$th diffusive mode is
supported on structures of characteristic size
\[
\ell_p(k) \;\sim\; \frac{\sqrt{A_p}}{k}.
\]
Diffusion supplies a competing length scale, $\ell_D \sim \sqrt{D}$, which sets the
smallest separation the operator can resolve.

Two ladders become eligible for hybridization when their associated geometric scales
become comparable within diffusive resolution.  For a fixed branch $(p,k_p)$ and the
period--$1$ ladder, we therefore monitor the scale mismatch
\[
\delta_p(k)
 \;=\;
\left|
\ell_1(k) - \ell_p(k_p)
\right|
 \;=\;
\left|
\frac{\sqrt{A_1}}{k} - \frac{\sqrt{A_p}}{k_p}
\right| .
\]
which is determined purely by island geometry and asymptotic scaling.  Local minima of
$\delta_p(k)$ identify those $k$ for which the two ladders achieve their closest geometric
commensurability.  These minima are sharp and isolated, and their ordering is fixed in
advance by geometry, independent of $D$.

\begin{figure}[t]
  \centering
  \includegraphics[width=0.85\linewidth]{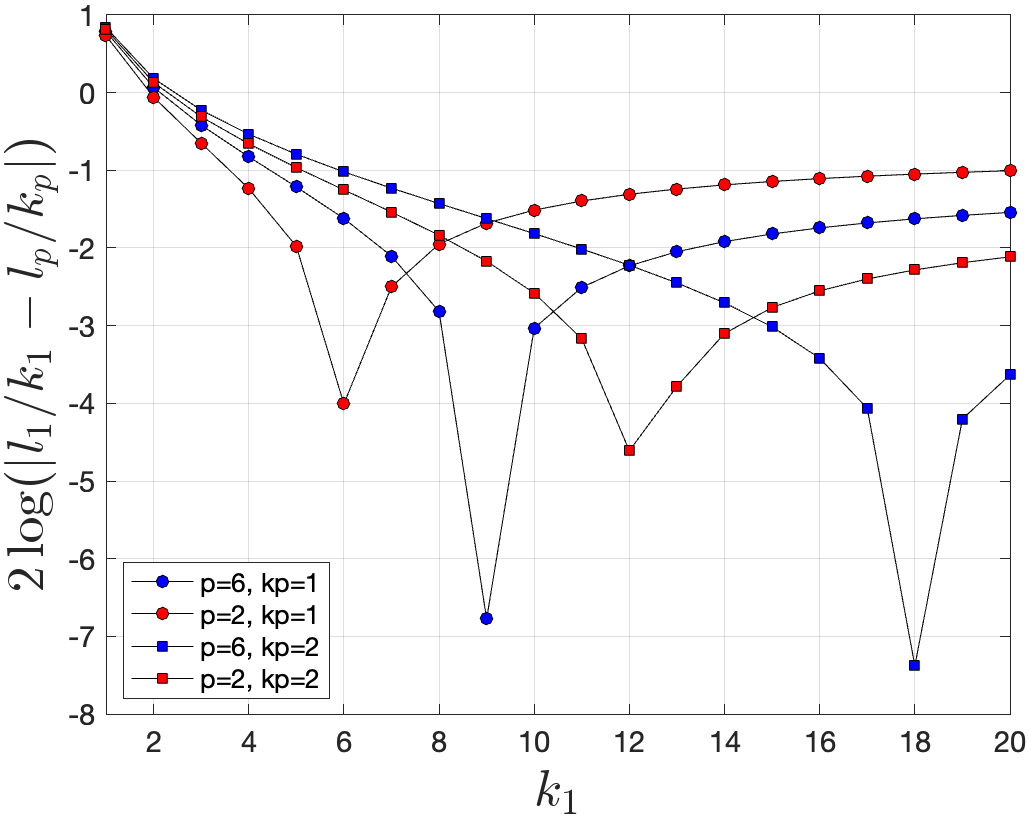}
  \caption{
  Detuning diagnostic used to predict candidate inter--island interactions.
  For each $k$ in the period--$1$ ladder we plot
  $2\log_{10}\!\left|\lambda^{(1)}_{k}/k - \lambda^{(p)}_{k_p}/k_p\right|$
  for $p\in\{2,6\}$ and selected $k_p$.
  }
  \label{fig:detuning_predictor}
\end{figure}

Diffusion controls whether a given geometric commensurability is actually realized as a
mixed eigenmode.  The operator cannot resolve scale separations smaller than
$\ell_D \sim \sqrt{D}$, so significant hybridization requires
\[
\delta_p(k) \;\lesssim\; \mathcal{O}(\sqrt{D}).
\]
At moderate $D$ several shallow minima satisfy this criterion, leading to early and
sometimes strong mixing between ladders.  As $D$ decreases, the resolvable threshold
tightens: many candidate crossings devolve into weak, percent--level leakage between
branches, while only the deepest geometric minima survive as genuine mixed modes.

Figure~\ref{fig:detuning_predictor} shows $\delta_p(k)$ for the period--$2$ and
period--$6$ branches against the period--$1$ ladder.  Each curve consists of a
sequence of sharp, isolated minima.  These minima mark the discrete values of $k$
for which the geometric length scale $\ell_1(k)$ most closely matches
$\ell_p(k_p)$ for a given island.  Their locations are fixed entirely by island
areas and ladder indices; they therefore provide a purely geometric ordering of
candidate inter--island interactions.

Diffusion does not determine where these minima occur.  It determines which of them
are dynamically relevant.  At moderate $D$, several of the shallow minima lie within
diffusive resolution and produce visibly mixed modes.  As $D$ decreases, the
resolvable threshold tightens: most of these candidates devolve into weak,
percent--level leakage, and only the deepest minima survive as genuine hybrid modes.
In the $D\!\to\!0$ limit, the first \emph{strong} mixed mode converges toward the
deepest geometric minimum, where the scale matching is most precise.

\subsection{Geometric predictors for the onset of advective modes}

A similar estimate governs the relative placement of diffusive and advective families.
For a fixed advective family (phase index) $q$, the first core--localized mode $\phi^{1}_{q,1}$ satisfies
\[
\gamma^{(\phi)}_{q,1} \;\sim\; \sqrt{D}\,\beta_q .
\]
Equating this with the $k$th diffusive mode of the period--$1$ island gives
\[
D\,\alpha_1 k^2 \;\approx\; \sqrt{D}\,\beta_q,
\qquad
k^*(q) \;\sim\; D^{-1/4}\,(\beta_q/\alpha_1)^{1/2}.
\]
Thus the number of diffusive modes lying below a given advective ladder grows as
$D^{-1/4}$ (equivalently $k^*\propto \mathrm{Pe}^{1/4}$), explaining quantitatively why
the Koopman ladders retreat to higher and higher indices as $\mathrm{Pe}$ increases.
Different advective families share the same exponent but may differ in prefactor
through the family--dependent constants $\beta_q$.

\begin{figure}[t]
\centering
\includegraphics[width=0.85\linewidth]{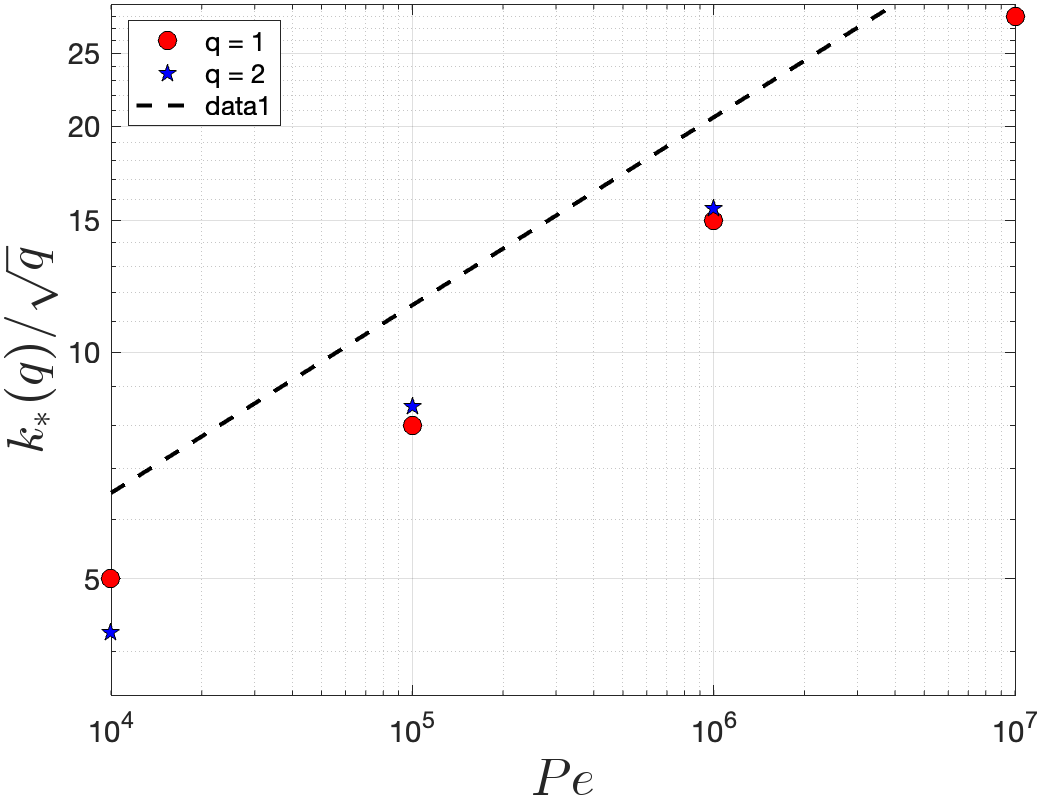}
\caption{
Normalized crossover index $k^*(q)/\sqrt{q}$ versus $\mathrm{Pe}$ for two advective families on the
period--$1$ island ($q=1$ circles, $q=2$ markers).
The dashed line indicates the predicted scaling $k^* \propto \mathrm{Pe}^{1/4}$, demonstrating collapse across $q$.
}
\label{fig:kstar_scaling}
\end{figure}

Figure~\ref{fig:kstar_scaling} verifies this prediction directly.
For each family we define $k^*(q)$ as the largest diffusive index whose decay rate
remains below that of the leading core--localized mode.
The measured crossover indices collapse onto straight lines in log--log coordinates
with slope $1/4$, in agreement with the $D^{-1/4}$ scaling.
The two families are approximately parallel but offset, indicating that the exponent
is universal while the prefactor depends on the local structure of the corresponding
elliptic region.

The estimates depend on local geometric information: island area, effective width, and 
curvature near elliptic points. No global spectral computation is required to determine 
the scaling of the crossover. From phase-space geometry alone, one may estimate which 
island families contribute the slowest diffusive modes, where advective families first 
enter the ordered spectrum, and when tunneling between branches occurs.

The mixed Hamiltonian geometry directly determines the organization of the 
advection-diffusion spectrum: which island families contribute the slowest modes, where 
advective families enter, and when hybridization occurs. The precise placement of 
individual modes remains sensitive to finite-$\mathrm{Pe}$ corrections and hybridization.

\section{Conclusions}

We have examined the spectral properties of the one-period advection-diffusion operator in 
a prototypical area-preserving map with mixed phase space. By combining direct numerical 
computations at Péclet numbers as large as $10^{7}$ with a semiclassical interpretation 
based on the effective parameter $\hbar_{\mathrm{eff}}=\sqrt{D}$, we have shown that a 
substantial portion of the slow spectrum admits a rigid geometric organization.

\section{Conclusions}

We have examined the spectral properties of the one-period advection-diffusion operator in 
a prototypical area-preserving map with mixed phase space. By combining direct numerical 
computations at Péclet numbers as large as $10^{7}$ with a semiclassical interpretation 
based on the effective parameter $\hbar_{\mathrm{eff}}=\sqrt{D}$, we have shown that a 
substantial portion of the slow spectrum admits a rigid geometric organization. Individual 
eigenmodes may be associated with specific phase-space structures: modes supported on 
entire islands behave as square-well states, modes localized near elliptic fixed points 
form harmonic-oscillator ladders, and additional diffusive modes arise from weak coupling 
between dynamically separated islands.

The semiclassical viewpoint provides not only a qualitative framework for interpreting 
these observations but also quantitative predictions for spectral scaling and multiplicity. 
In particular, the number of diffusive modes supported on a given island and their 
decay-rate scaling follow directly from the size of the corresponding invariant set and the 
dependence of $\hbar_{\mathrm{eff}}$ on $D$. The close agreement between these predictions 
and the numerical results confirms that the observed organization is controlled by the 
underlying Lagrangian geometry rather than by accidental features of discretization or 
parameter choice. While individual scaling laws have been established for limiting 
cases—$\mathrm{Pe}^{-1/2}$ for integrable flows~\cite{Vukadinovic2015,bedrossian2017} and 
$\mathrm{Pe}^{-1}$ for dominant island modes~\cite{Gorodetskyi:2012PoFa,Gorodetskyi:2012PoFb}—the 
present work demonstrates how these families coexist and compete across the full slow 
spectrum.

Inter-island coupling plays a central role in shaping the slow spectrum. At finite 
diffusivity we observe mixed eigenfunctions whose supports span multiple dynamically 
separated islands. Our geometric resonance model shows that this hybridization is not a 
monotone finite-diffusivity effect that simply disappears as $D\to 0$; instead, it is 
selected by near-commensurabilities between island areas and local length scales, and 
therefore can recur at moderate values of the radial index $k$ even at very small $D$. 
This provides a geometric explanation for the near-degeneracies observed by Popovych et 
al.~\cite{Popovych2007}, which were found to vary unpredictably in parameter space. Mixed 
diffusive structures thus persist in the slow spectrum in a geometry-controlled way.

As Péclet number increases, the advective (Koopman) modes retreat to higher and higher 
indices, consistent with the $\mathrm{Pe}^{1/4}$ scaling derived in 
Sec.~\ref{sec:prediction}. As a result, for any fixed mode count N, the first N 
eigenmodes become increasingly dominated by diffusive, square-well-like island families. 
In this sense, finite-mode spectral approximations become effectively self-adjoint as 
$\mathrm{Pe}\to\infty$: the leading modes are increasingly well described by self-adjoint 
diffusive operators on individual regular sets.

This reorganization by mode number, however, does not imply that advective modes become 
dynamically irrelevant. For modes slower than a fixed decay rate $\gamma^*$—the 
physically meaningful criterion for finite-time dynamics—both diffusive and advective 
families contribute $O(\sqrt{\mathrm{Pe}})$ modes, with their ratio determined by 
geometric prefactors independent of Pe. The $\mathrm{Pe}^{1/4}$ crossover reorganizes 
where families appear in the ordered spectrum but does not change their relative 
importance for finite-time dynamics. The practical consequence is that spectral 
approximations based on fixed mode counts (common in reduced-order models) capture 
increasingly diffusive character, while approximations based on fixed timescales retain 
balanced contributions from both families.

An important consequence of this structure is the absence of any a priori geometric 
control over spectral gaps within the slow manifold. While the underlying phase-space 
geometry predicts where competing diffusive ladders approach one another, the resulting 
interactions generically produce avoided crossings whose detailed splittings depend on 
fine-scale structure. As a result, the ordering and separation of decay rates within the 
slow spectrum are not monotone functions of island size or resonance order. Mixed modes 
therefore disrupt any simple hierarchy of time scales, and preclude a universal geometric 
ranking of finite-time decay channels.

These results show that the slow decay of passive scalars in mixed phase spaces is 
governed neither by a uniform chaotic mechanism nor by a single dominant eigenmode, but by 
a semiclassical partitioning of phase space. Invariant sets of the Lagrangian dynamics 
imprint themselves directly on the spectrum, giving rise to multiple long-lived decay 
channels whose scaling, multiplicity, and ordering are fixed by geometry but modulated by 
resonance. Mixed phase spaces therefore admit a natural spectral language for transport, 
in which Lagrangian structure determines not only where scalars persist, but how the 
entire slow spectrum is organized.

The analysis presented here suggests several directions for future work, including 
extensions to continuously driven or aperiodic flows, the role of island breakup and 
transport barriers in higher dimensions, and the impact of stochastic perturbations on 
semiclassical scaling. More broadly, the emergence of geometry-controlled spectral 
families highlights mixed Hamiltonian systems as a natural setting in which 
quantum-inspired ideas can be fruitfully applied to classical transport problems.

\bibliography{AdvDiff}

%aipnum4-2.bst 2019-01-14 (MD) hand-edited version of apsrev4-1.bst
\begin{thebibliography}{23}%
\makeatletter
\providecommand \@ifxundefined [1]{%
 \@ifx{#1\undefined}
}%
\providecommand \@ifnum [1]{%
 \ifnum #1\expandafter \@firstoftwo
 \else \expandafter \@secondoftwo
 \fi
}%
\providecommand \@ifx [1]{%
 \ifx #1\expandafter \@firstoftwo
 \else \expandafter \@secondoftwo
 \fi
}%
\providecommand \natexlab [1]{#1}%
\providecommand \enquote  [1]{``#1''}%
\providecommand \bibnamefont  [1]{#1}%
\providecommand \bibfnamefont [1]{#1}%
\providecommand \citenamefont [1]{#1}%
\providecommand \href@noop [0]{\@secondoftwo}%
\providecommand \href [0]{\begingroup \@sanitize@url \@href}%
\providecommand \@href[1]{\@@startlink{#1}\@@href}%
\providecommand \@@href[1]{\endgroup#1\@@endlink}%
\providecommand \@sanitize@url [0]{\catcode `\\12\catcode `\$12\catcode
  `\&12\catcode `\#12\catcode `\^12\catcode `\_12\catcode `\%12\relax}%
\providecommand \@@startlink[1]{}%
\providecommand \@@endlink[0]{}%
\providecommand \url  [0]{\begingroup\@sanitize@url \@url }%
\providecommand \@url [1]{\endgroup\@href {#1}{\urlprefix }}%
\providecommand \urlprefix  [0]{URL }%
\providecommand \Eprint [0]{\href }%
\providecommand \doibase [0]{https://doi.org/}%
\providecommand \selectlanguage [0]{\@gobble}%
\providecommand \bibinfo  [0]{\@secondoftwo}%
\providecommand \bibfield  [0]{\@secondoftwo}%
\providecommand \translation [1]{[#1]}%
\providecommand \BibitemOpen [0]{}%
\providecommand \bibitemStop [0]{}%
\providecommand \bibitemNoStop [0]{.\EOS\space}%
\providecommand \EOS [0]{\spacefactor3000\relax}%
\providecommand \BibitemShut  [1]{\csname bibitem#1\endcsname}%
\let\auto@bib@innerbib\@empty
%</preamble>
\bibitem [{\citenamefont {Aref}\ \emph {et~al.}(2017)\citenamefont {Aref} \emph
  {et~al.}}]{Aref2017}%
  \BibitemOpen
  \bibfield  {author} {\bibinfo {author} {\bibfnamefont {H.}~\bibnamefont
  {Aref}} \emph {et~al.},\ }\bibfield  {title} {\enquote {\bibinfo {title}
  {Frontiers of chaotic advection},}\ }\href
  {https://doi.org/10.1103/RevModPhys.89.025007} {\bibfield  {journal}
  {\bibinfo  {journal} {Reviews of Modern Physics}\ }\textbf {\bibinfo {volume}
  {89}},\ \bibinfo {pages} {025007} (\bibinfo {year} {2017})}\BibitemShut
  {NoStop}%
\bibitem [{\citenamefont {Babiano}\ \emph {et~al.}(1994)\citenamefont
  {Babiano}, \citenamefont {Boffetta}, \citenamefont {Provenzale},\ and\
  \citenamefont {Vulpiani}}]{Babiano1994}%
  \BibitemOpen
  \bibfield  {author} {\bibinfo {author} {\bibfnamefont {A.}~\bibnamefont
  {Babiano}}, \bibinfo {author} {\bibfnamefont {G.}~\bibnamefont {Boffetta}},
  \bibinfo {author} {\bibfnamefont {A.}~\bibnamefont {Provenzale}},\ and\
  \bibinfo {author} {\bibfnamefont {A.}~\bibnamefont {Vulpiani}},\ }\bibfield
  {title} {\enquote {\bibinfo {title} {Chaotic advection in point vortex models
  and two-dimensional turbulence},}\ }\href@noop {} {\bibfield  {journal}
  {\bibinfo  {journal} {Physics of Fluids}\ }\textbf {\bibinfo {volume} {6}},\
  \bibinfo {pages} {2465--2474} (\bibinfo {year} {1994})}\BibitemShut {NoStop}%
\bibitem [{\citenamefont {Benzekri}\ \emph {et~al.}(2006)\citenamefont
  {Benzekri}, \citenamefont {Chandre}, \citenamefont {Leoncini}, \citenamefont
  {Lima},\ and\ \citenamefont {Vittot}}]{Benzekri2006}%
  \BibitemOpen
  \bibfield  {author} {\bibinfo {author} {\bibfnamefont {T.}~\bibnamefont
  {Benzekri}}, \bibinfo {author} {\bibfnamefont {C.}~\bibnamefont {Chandre}},
  \bibinfo {author} {\bibfnamefont {X.}~\bibnamefont {Leoncini}}, \bibinfo
  {author} {\bibfnamefont {R.}~\bibnamefont {Lima}},\ and\ \bibinfo {author}
  {\bibfnamefont {M.}~\bibnamefont {Vittot}},\ }\bibfield  {title} {\enquote
  {\bibinfo {title} {Chaotic advection and targeted mixing},}\ }\href@noop {}
  {\bibfield  {journal} {\bibinfo  {journal} {Physical Review Letters}\
  }\textbf {\bibinfo {volume} {96}} (\bibinfo {year} {2006})}\BibitemShut
  {NoStop}%
\bibitem [{\citenamefont {Schlick}, \citenamefont {Lueptow},\ and\
  \citenamefont {Dorfman}(2013)}]{Schlick2013}%
  \BibitemOpen
  \bibfield  {author} {\bibinfo {author} {\bibfnamefont {C.}~\bibnamefont
  {Schlick}}, \bibinfo {author} {\bibfnamefont {R.~M.}\ \bibnamefont
  {Lueptow}},\ and\ \bibinfo {author} {\bibfnamefont {K.~D.}\ \bibnamefont
  {Dorfman}},\ }\bibfield  {title} {\enquote {\bibinfo {title} {Interplay
  between chaos and diffusion in time-periodic sine flow},}\ }\href
  {https://doi.org/10.1063/1.4824732} {\bibfield  {journal} {\bibinfo
  {journal} {Physics of Fluids}\ }\textbf {\bibinfo {volume} {25}},\ \bibinfo
  {pages} {103602} (\bibinfo {year} {2013})}\BibitemShut {NoStop}%
\bibitem [{\citenamefont {Lopez}\ \emph {et~al.}(2001)\citenamefont {Lopez},
  \citenamefont {Neufeld}, \citenamefont {Hernandez-Garcia},\ and\
  \citenamefont {Haynes}}]{Lopez2001}%
  \BibitemOpen
  \bibfield  {author} {\bibinfo {author} {\bibfnamefont {C.}~\bibnamefont
  {Lopez}}, \bibinfo {author} {\bibfnamefont {Z.}~\bibnamefont {Neufeld}},
  \bibinfo {author} {\bibfnamefont {E.}~\bibnamefont {Hernandez-Garcia}},\ and\
  \bibinfo {author} {\bibfnamefont {P.~H.}\ \bibnamefont {Haynes}},\ }\bibfield
   {title} {\enquote {\bibinfo {title} {Chaotic advection of reacting
  substances: Plankton dynamics on a meandering jet},}\ }\href@noop {}
  {\bibfield  {journal} {\bibinfo  {journal} {Physics and Chemistry of the
  Earth Part {B}: Hydrology, Oceans and Atmosphere}\ }\textbf {\bibinfo
  {volume} {26}},\ \bibinfo {pages} {313--317} (\bibinfo {year}
  {2001})}\BibitemShut {NoStop}%
\bibitem [{\citenamefont {Prants}\ \emph {et~al.}(2011)\citenamefont {Prants},
  \citenamefont {Budyansky}, \citenamefont {Ponomarev},\ and\ \citenamefont
  {Uleysky}}]{Prants2011}%
  \BibitemOpen
  \bibfield  {author} {\bibinfo {author} {\bibfnamefont {S.}~\bibnamefont
  {Prants}}, \bibinfo {author} {\bibfnamefont {M.}~\bibnamefont {Budyansky}},
  \bibinfo {author} {\bibfnamefont {V.}~\bibnamefont {Ponomarev}},\ and\
  \bibinfo {author} {\bibfnamefont {M.}~\bibnamefont {Uleysky}},\ }\bibfield
  {title} {\enquote {\bibinfo {title} {Lagrangian study of transport and mixing
  in a mesoscale eddy street},}\ }\href
  {https://doi.org/10.1016/j.ocemod.2011.02.008} {\bibfield  {journal}
  {\bibinfo  {journal} {Ocean Modelling}\ }\textbf {\bibinfo {volume} {38}},\
  \bibinfo {pages} {114--125} (\bibinfo {year} {2011})}\BibitemShut {NoStop}%
\bibitem [{\citenamefont {Elgindi}, \citenamefont {Liss},\ and\ \citenamefont
  {Mattingly}(2025)}]{elgindi2025optimal}%
  \BibitemOpen
  \bibfield  {author} {\bibinfo {author} {\bibfnamefont {T.~M.}\ \bibnamefont
  {Elgindi}}, \bibinfo {author} {\bibfnamefont {K.}~\bibnamefont {Liss}},\ and\
  \bibinfo {author} {\bibfnamefont {J.~C.}\ \bibnamefont {Mattingly}},\
  }\bibfield  {title} {\enquote {\bibinfo {title} {Optimal enhanced dissipation
  and mixing for a time-periodic, lipschitz velocity field on
  $\mathbb{T}^2$},}\ }\href {https://doi.org/10.1215/00127094-2024-0057}
  {\bibfield  {journal} {\bibinfo  {journal} {Duke Mathematical Journal}\
  }\textbf {\bibinfo {volume} {174}},\ \bibinfo {pages} {1209--1260} (\bibinfo
  {year} {2025})}\BibitemShut {NoStop}%
\bibitem [{\citenamefont {Pierrehumbert}(1994)}]{Pierrehumbert:1994}%
  \BibitemOpen
  \bibfield  {author} {\bibinfo {author} {\bibfnamefont {R.~T.}\ \bibnamefont
  {Pierrehumbert}},\ }\bibfield  {title} {\enquote {\bibinfo {title} {Tracer
  microstructure in the large-eddy dominated regime},}\ }\href@noop {}
  {\bibfield  {journal} {\bibinfo  {journal} {Chaos Solitons \& Fractals}\
  }\textbf {\bibinfo {volume} {4}},\ \bibinfo {pages} {1091--1110} (\bibinfo
  {year} {1994})}\BibitemShut {NoStop}%
\bibitem [{\citenamefont {Toussaint}\ \emph {et~al.}(2000)\citenamefont
  {Toussaint}, \citenamefont {Carrière}, \citenamefont {Scott},\ and\
  \citenamefont {Gence}}]{Toussaint2000}%
  \BibitemOpen
  \bibfield  {author} {\bibinfo {author} {\bibfnamefont {V.}~\bibnamefont
  {Toussaint}}, \bibinfo {author} {\bibfnamefont {P.}~\bibnamefont
  {Carrière}}, \bibinfo {author} {\bibfnamefont {J.}~\bibnamefont {Scott}},\
  and\ \bibinfo {author} {\bibfnamefont {J.-N.}\ \bibnamefont {Gence}},\
  }\bibfield  {title} {\enquote {\bibinfo {title} {Spectral decay of a passive
  scalar in chaotic mixing},}\ }\href
  {https://doi.org/http://dx.doi.org/10.1063/1.1290277} {\bibfield  {journal}
  {\bibinfo  {journal} {Physics of Fluids (1994-present)}\ }\textbf {\bibinfo
  {volume} {12}},\ \bibinfo {pages} {2834--2844} (\bibinfo {year}
  {2000})}\BibitemShut {NoStop}%
\bibitem [{\citenamefont {Sundaram}, \citenamefont {Poje},\ and\ \citenamefont
  {Pattanayak}(2009)}]{Sundaram:2009PRE}%
  \BibitemOpen
  \bibfield  {author} {\bibinfo {author} {\bibfnamefont {B.}~\bibnamefont
  {Sundaram}}, \bibinfo {author} {\bibfnamefont {A.~C.}\ \bibnamefont {Poje}},\
  and\ \bibinfo {author} {\bibfnamefont {A.~K.}\ \bibnamefont {Pattanayak}},\
  }\bibfield  {title} {\enquote {\bibinfo {title} {Persistent patterns and
  multifractality in fluid mixing},}\ }\href
  {https://doi.org/10.1103/PhysRevE.79.066202} {\bibfield  {journal} {\bibinfo
  {journal} {Phys. Rev. E}\ }\textbf {\bibinfo {volume} {79}},\ \bibinfo
  {pages} {066202} (\bibinfo {year} {2009})}\BibitemShut {NoStop}%
\bibitem [{\citenamefont {Popovych}, \citenamefont {Pikovsky},\ and\
  \citenamefont {Eckhardt}(2007)}]{Popovych2007}%
  \BibitemOpen
  \bibfield  {author} {\bibinfo {author} {\bibfnamefont {O.~V.}\ \bibnamefont
  {Popovych}}, \bibinfo {author} {\bibfnamefont {A.}~\bibnamefont {Pikovsky}},\
  and\ \bibinfo {author} {\bibfnamefont {B.}~\bibnamefont {Eckhardt}},\
  }\bibfield  {title} {\enquote {\bibinfo {title} {Abnormal mixing of passive
  scalars in chaotic flows},}\ }\href
  {https://doi.org/10.1103/PhysRevE.75.036308} {\bibfield  {journal} {\bibinfo
  {journal} {Physical Review E}\ }\textbf {\bibinfo {volume} {75}},\ \bibinfo
  {pages} {036308} (\bibinfo {year} {2007})}\BibitemShut {NoStop}%
\bibitem [{\citenamefont {Mezić}(2005)}]{Mezic2005}%
  \BibitemOpen
  \bibfield  {author} {\bibinfo {author} {\bibfnamefont {I.}~\bibnamefont
  {Mezić}},\ }\bibfield  {title} {\enquote {\bibinfo {title} {Spectral
  properties of dynamical systems, model reduction and decompositions},}\
  }\href {https://doi.org/10.1007/s11071-005-2824-x} {\bibfield  {journal}
  {\bibinfo  {journal} {Nonlinear Dynamics}\ }\textbf {\bibinfo {volume}
  {41}},\ \bibinfo {pages} {309--325} (\bibinfo {year} {2005})}\BibitemShut
  {NoStop}%
\bibitem [{\citenamefont {Mezić}(2013)}]{Mezic2013}%
  \BibitemOpen
  \bibfield  {author} {\bibinfo {author} {\bibfnamefont {I.}~\bibnamefont
  {Mezić}},\ }\bibfield  {title} {\enquote {\bibinfo {title} {Analysis of
  fluid flows via spectral properties of the koopman operator},}\ }\href
  {https://doi.org/10.1146/annurev-fluid-011212-140652} {\bibfield  {journal}
  {\bibinfo  {journal} {Annual Review of Fluid Mechanics}\ }\textbf {\bibinfo
  {volume} {45}},\ \bibinfo {pages} {357--378} (\bibinfo {year}
  {2013})}\BibitemShut {NoStop}%
\bibitem [{\citenamefont {Froyland}(2010)}]{Froyland2010}%
  \BibitemOpen
  \bibfield  {author} {\bibinfo {author} {\bibfnamefont {G.}~\bibnamefont
  {Froyland}},\ }\bibfield  {title} {\enquote {\bibinfo {title} {Transport in
  dynamical systems: coherent sets},}\ }\href
  {https://doi.org/10.1007/s00220-006-0030-0} {\bibfield  {journal} {\bibinfo
  {journal} {Communications in Mathematical Physics}\ }\textbf {\bibinfo
  {volume} {268}},\ \bibinfo {pages} {413--449} (\bibinfo {year}
  {2010})}\BibitemShut {NoStop}%
\bibitem [{\citenamefont {Giona}, \citenamefont {Cerbelli},\ and\ \citenamefont
  {Adrover}(2004)}]{Giona2004}%
  \BibitemOpen
  \bibfield  {author} {\bibinfo {author} {\bibfnamefont {M.}~\bibnamefont
  {Giona}}, \bibinfo {author} {\bibfnamefont {S.}~\bibnamefont {Cerbelli}},\
  and\ \bibinfo {author} {\bibfnamefont {A.}~\bibnamefont {Adrover}},\
  }\bibfield  {title} {\enquote {\bibinfo {title} {Spectral properties and
  transport mechanisms of partially chaotic bounded flows in the presence of
  diffusion},}\ }\href {https://doi.org/10.1063/1.1626115} {\bibfield
  {journal} {\bibinfo  {journal} {Physics of Fluids}\ }\textbf {\bibinfo
  {volume} {16}},\ \bibinfo {pages} {141--154} (\bibinfo {year}
  {2004})}\BibitemShut {NoStop}%
\bibitem [{\citenamefont {Cerbelli}\ \emph {et~al.}(2004)\citenamefont
  {Cerbelli}, \citenamefont {Vitacolonna}, \citenamefont {Adrover},\ and\
  \citenamefont {Giona}}]{Cerbelli:2004}%
  \BibitemOpen
  \bibfield  {author} {\bibinfo {author} {\bibfnamefont {S.}~\bibnamefont
  {Cerbelli}}, \bibinfo {author} {\bibfnamefont {V.}~\bibnamefont
  {Vitacolonna}}, \bibinfo {author} {\bibfnamefont {A.}~\bibnamefont
  {Adrover}},\ and\ \bibinfo {author} {\bibfnamefont {M.}~\bibnamefont
  {Giona}},\ }\bibfield  {title} {\enquote {\bibinfo {title}
  {Eigenvalue-eigenfunction analysis of infinitely fast reactions and
  micromixing regimes in regular and chaotic bounded flows},}\ }\href@noop {}
  {\bibfield  {journal} {\bibinfo  {journal} {Chem. Eng. Sci.}\ }\textbf
  {\bibinfo {volume} {59}},\ \bibinfo {pages} {2125--2144} (\bibinfo {year}
  {2004})}\BibitemShut {NoStop}%
\bibitem [{\citenamefont {Gorodetskyi}, \citenamefont {Speetjens},\ and\
  \citenamefont {Anderson}(2012)}]{Gorodetskyi:2012PoFa}%
  \BibitemOpen
  \bibfield  {author} {\bibinfo {author} {\bibfnamefont {O.}~\bibnamefont
  {Gorodetskyi}}, \bibinfo {author} {\bibfnamefont {M.~F.~M.}\ \bibnamefont
  {Speetjens}},\ and\ \bibinfo {author} {\bibfnamefont {P.~D.}\ \bibnamefont
  {Anderson}},\ }\bibfield  {title} {\enquote {\bibinfo {title} {An efficient
  approach for eigenmode analysis of transient distributive mixing by the
  mapping method},}\ }\href {https://doi.org/10.1063/1.4712133} {\bibfield
  {journal} {\bibinfo  {journal} {Physics of Fluids}\ }\textbf {\bibinfo
  {volume} {24}},\ \bibinfo {eid} {053602} (\bibinfo {year}
  {2012})}\BibitemShut {NoStop}%
\bibitem [{\citenamefont {Gorodetskyi}, \citenamefont {Giona},\ and\
  \citenamefont {Anderson}(2012)}]{Gorodetskyi:2012PoFb}%
  \BibitemOpen
  \bibfield  {author} {\bibinfo {author} {\bibfnamefont {O.}~\bibnamefont
  {Gorodetskyi}}, \bibinfo {author} {\bibfnamefont {M.}~\bibnamefont {Giona}},\
  and\ \bibinfo {author} {\bibfnamefont {P.~D.}\ \bibnamefont {Anderson}},\
  }\bibfield  {title} {\enquote {\bibinfo {title} {Spectral analysis of mixing
  in chaotic flows via the mapping matrix formalism: Inclusion of molecular
  diffusion and quantitative eigenvalue estimate in the purely convective
  limit},}\ }\href {https://doi.org/10.1063/1.4738598} {\bibfield  {journal}
  {\bibinfo  {journal} {Physics of Fluids}\ }\textbf {\bibinfo {volume} {24}},\
  \bibinfo {eid} {073603} (\bibinfo {year} {2012})}\BibitemShut {NoStop}%
\bibitem [{\citenamefont {Vukadinovic}\ \emph {et~al.}(2015)\citenamefont
  {Vukadinovic}, \citenamefont {Dedits}, \citenamefont {Poje},\ and\
  \citenamefont {Sch{\"a}fer}}]{Vukadinovic2015}%
  \BibitemOpen
  \bibfield  {author} {\bibinfo {author} {\bibfnamefont {J.}~\bibnamefont
  {Vukadinovic}}, \bibinfo {author} {\bibfnamefont {E.}~\bibnamefont {Dedits}},
  \bibinfo {author} {\bibfnamefont {A.~C.}\ \bibnamefont {Poje}},\ and\
  \bibinfo {author} {\bibfnamefont {T.}~\bibnamefont {Sch{\"a}fer}},\
  }\bibfield  {title} {\enquote {\bibinfo {title} {Averaging and spectral
  properties for the 2d advection--diffusion equation in the semiclassical
  limit},}\ }\href {https://doi.org/10.1016/j.physd.2015.08.001} {\bibfield
  {journal} {\bibinfo  {journal} {Physica D}\ }\textbf {\bibinfo {volume}
  {310}},\ \bibinfo {pages} {1--18} (\bibinfo {year} {2015})}\BibitemShut
  {NoStop}%
\bibitem [{\citenamefont {Bedrossian}\ and\ \citenamefont
  {Coti~Zelati}(2017)}]{bedrossian2017}%
  \BibitemOpen
  \bibfield  {author} {\bibinfo {author} {\bibfnamefont {J.}~\bibnamefont
  {Bedrossian}}\ and\ \bibinfo {author} {\bibfnamefont {M.}~\bibnamefont
  {Coti~Zelati}},\ }\bibfield  {title} {\enquote {\bibinfo {title} {Enhanced
  dissipation, hypoellipticity, and anomalous small noise inviscid limits in
  shear flows},}\ }\href@noop {} {\bibfield  {journal} {\bibinfo  {journal}
  {Archive for Rational Mechanics and Analysis}\ }\textbf {\bibinfo {volume}
  {224}},\ \bibinfo {pages} {1161--1204} (\bibinfo {year} {2017})}\BibitemShut
  {NoStop}%
\bibitem [{\citenamefont {Vukadinovic}(2021)}]{Vukadinovic2021}%
  \BibitemOpen
  \bibfield  {author} {\bibinfo {author} {\bibfnamefont {J.}~\bibnamefont
  {Vukadinovic}},\ }\bibfield  {title} {\enquote {\bibinfo {title} {The limit
  of vanishing diffusivity for passive scalars in hamiltonian flows},}\ }\href
  {https://doi.org/10.1007/s00205-021-01707-7} {\bibfield  {journal} {\bibinfo
  {journal} {Archive for Rational Mechanics and Analysis}\ }\textbf {\bibinfo
  {volume} {242}},\ \bibinfo {pages} {1395--1444} (\bibinfo {year}
  {2021})}\BibitemShut {NoStop}%
\bibitem [{\citenamefont {Chirikov}(1979)}]{Chirikov1979}%
  \BibitemOpen
  \bibfield  {author} {\bibinfo {author} {\bibfnamefont {B.~V.}\ \bibnamefont
  {Chirikov}},\ }\bibfield  {title} {\enquote {\bibinfo {title} {A universal
  instability of many-dimensional oscillator systems},}\ }\href@noop {}
  {\bibfield  {journal} {\bibinfo  {journal} {Physics Reports}\ }\textbf
  {\bibinfo {volume} {52}},\ \bibinfo {pages} {263--379} (\bibinfo {year}
  {1979})}\BibitemShut {NoStop}%
\bibitem [{\citenamefont {Lehoucq}, \citenamefont {Sorensen},\ and\
  \citenamefont {Yang}(1998)}]{arpack1998}%
  \BibitemOpen
  \bibfield  {author} {\bibinfo {author} {\bibfnamefont {R.~B.}\ \bibnamefont
  {Lehoucq}}, \bibinfo {author} {\bibfnamefont {D.~C.}\ \bibnamefont
  {Sorensen}},\ and\ \bibinfo {author} {\bibfnamefont {C.}~\bibnamefont
  {Yang}},\ }\href@noop {} {\emph {\bibinfo {title} {ARPACK Users' Guide:
  Solution of Large-Scale Eigenvalue Problems with Implicitly Restarted Arnoldi
  Methods}}}\ (\bibinfo  {publisher} {Society for Industrial and Applied
  Mathematics},\ \bibinfo {address} {Philadelphia, PA},\ \bibinfo {year}
  {1998})\BibitemShut {NoStop}%
\end{thebibliography}%

\appendix

\section*{Appendix A: Numerical Implementation}\label{app:numerics}

The leading eigenvalues and eigenfunctions of the advection-diffusion operator are 
computed using an implicitly restarted Arnoldi method as implemented in ARPACK~\cite{arpack1998}. The 
operator is never formed explicitly. Instead, ARPACK is supplied with a matrix-vector 
product routine that applies one step of the Fourier-space transfer operator to an input 
vector. All iterations therefore operate in a matrix-free setting, with the computational 
cost dominated by repeated evaluations of this operator.

The state is represented in Fourier space on a square spectral grid
\[
(m,n)\in\{-l,\ldots,l\}^2,
\]
so that the dimension of the eigenproblem is
\[
N = (2l+1)^2 .
\]
The integer $l$ is read from an external parameter file and directly controls the spatial 
resolution. When eigenfunctions are reconstructed in physical space, they are interpreted 
on an $N_x\times N_x$ grid with $N_x=2l+1$ points in each direction, corresponding to a 
periodic domain of size $2\pi\times 2\pi$ and grid spacing $\Delta x=\Delta y=2\pi/(2l+1)$.

The matrix-vector product implements the one-step advection-diffusion operator in Fourier 
space,
\[
\rho^{t+1}_{m,n}
=
e^{-D(m^2+n^2)}
\sum_{k} J_{m-k}(n\kappa)\,\rho^{t}_{k,k+n},
\]
with truncation-aware bounds on the summation index $k$ to ensure that all accessed modes 
remain within the spectral domain. Here $J_q$ denotes the Bessel function of the first 
kind, $\kappa$ is the map amplitude, and $D$ is the diffusion coefficient. The Bessel 
weights $J_q(n\kappa)$ are precomputed once at startup. The operator is applied by direct 
convolution and index shifting in coefficient space, and is parallelized with OpenMP over 
Fourier modes.

The Arnoldi iteration targets the eigenvalues of largest magnitude, corresponding to the 
least damped modes of the transfer operator. Convergence is assessed using the standard 
relative residual
\[
\frac{\|A v - \lambda v\|_2}{|\lambda|}.
\]
Typical runs achieve residuals of order $10^{-14}$, consistent with double-precision 
accuracy. The total cost of a computation is well characterized by the number of operator 
applications reported by ARPACK, with each application corresponding to one evaluation of 
the Fourier-space advection-diffusion operator.

As a representative high-resolution case, we consider $\kappa=1.7$ and $l=600$, for which 
the eigenproblem has dimension
\[
N = (2l+1)^2 = 1201^2 \approx 1.44\times 10^6,
\]
and the reconstructed eigenfunctions are resolved on a $1201\times1201$ physical grid with 
spacing $\Delta x = 2\pi/1201 \approx 7.8\times 10^{-3}$. For this problem, even at very 
weak diffusion ($D=10^{-7}$), the Arnoldi iteration converges to all requested eigenpairs 
with relative residuals
\[
\frac{\|A v - \lambda v\|_2}{|\lambda|} \lesssim 10^{-12},
\]
with the majority of modes achieving residuals at the $10^{-13}$--$10^{-14}$ level. These 
values are far below any physically meaningful scale associated with the discretized 
operator and indicate that the computed eigenpairs are accurate to essentially machine 
precision. The computational effort in this regime corresponds to $\mathcal{O}(10^3)$ 
applications of the Fourier-space operator.

\subsection*{Spatial Resolution and Spectral Adequacy}

The choice of truncation $l=600$ is governed by the exponential damping inherent to the 
diffusive operator. The factor $e^{-D(m^2+n^2)}$ in the Fourier-space evolution operator 
suppresses contributions from high wavenumbers at a rate that depends directly on the 
diffusivity. At the weakest diffusion considered here ($D=10^{-7}$), Fourier modes with 
$|m|$ or $|n|$ exceeding
\[
k_c \approx \sqrt{\frac{-\ln\epsilon}{D}} \approx \sqrt{\frac{32}{D}} \approx 566
\]
contribute less than $\epsilon = 10^{-14}$ to the operator in spectral norm, and are 
therefore below the numerical precision of the computation. The truncation at $l=600$ thus 
captures all dynamically significant modes, with a safety margin of approximately six 
percent.

Spectral convergence is further supported by three independent consistency checks. First, 
the tight Arnoldi residuals reported above confirm that the computed eigenpairs are 
converged to machine precision within the chosen Fourier basis. Second, the collapse of 
rescaled eigenvalue families onto universal curves across three decades in Péclet number 
(Figs.~9, 11, 13) demonstrates that the computation has reached the asymptotic 
weak-diffusion regime in which the predicted scaling laws hold. Systematic 
under-resolution would manifest as a drift or scatter in these scalings; none is observed. 
Third, the ordering and multiplicity of spectral families matches the predictions derived 
from phase-space geometry in Section~VII, providing an independent geometric validation of 
the numerical results. Additionally, Fourier truncation strategies at comparable or 
coarser resolution have been validated for related advection-diffusion problems with mixed 
phase spaces \cite{Giona2004,Gorodetskyi:2012PoFa,Gorodetskyi:2012PoFb}, where spectral 
convergence and modal structure were shown to be robust to discretization parameters in 
the weak-diffusion limit.

These diagnostics, taken together with the exponential filtering built into the operator 
itself, provide strong evidence that the reported spectra are fully resolved and that the 
observed organization into diffusive, advective, and tunneling families is a robust 
feature of the continuum problem rather than an artifact of discretization.

\section*{Appendix B: Local Hamiltonian Geometry of Regular Islands}
\label{app:heff}

The interpretation of the slow spectrum in terms of square-well and oscillator families 
relies on geometric properties of the regular islands of the underlying Lagrangian map. 
For each elliptic island we construct a local effective Hamiltonian 
$H_{\mathrm{eff}}(x,y)$ whose level sets coincide with invariant curves of the map. This 
provides two geometric parameters that govern the slow spectrum: an island scale $\alpha$, 
proportional to the square root of the island area, which controls the spacing of 
diffusive modes, and a core curvature $\beta$, the quadratic coefficient of 
$H_{\mathrm{eff}}$ at the elliptic fixed point, which controls the spacing of advective 
modes. We document how these quantities are obtained directly from the map, without 
reference to any spectral computation.

\subsection*{Reconstructing $H_{\mathrm{eff}}$}

Let $(x_\ast,y_\ast)$ be an elliptic fixed point of the map. We initialize trajectories 
at small offsets $(x_\ast+r,y_\ast)$ and iterate the map for $N\gg1$ steps, discarding an 
initial transient. The angular advance per iterate yields a local rotation number 
$\rho(r)$. Integrating this profile defines an effective action variable
\begin{equation}
I(r) = \int_0^r \rho(s)\, ds,
\end{equation}
and hence an effective Hamiltonian $H_{\mathrm{eff}}(I)$ satisfying 
$\Omega(I) = dH_{\mathrm{eff}}/dI = \rho$. Mapping this relation back into physical space 
produces $H_{\mathrm{eff}}(x,y)$ whose contours coincide with invariant curves of the 
island. Figure~\ref{fig:phase_portrait} shows reconstructed $H_{\mathrm{eff}}$ for the 
period-1 and period-6 islands at $K=1.7$.

\subsection*{Island scale and diffusive modes}

The outer boundary of each island forms a finite barrier for diffusion. The effective 
well width $\alpha_p$ is defined by
\begin{equation}
\alpha_p^2 \equiv \frac{1}{\pi} \mathrm{Area}(\text{island } p),
\end{equation}
so that $\alpha_p$ is the radius of a disk with the same area. Diffusion inside an island 
is approximately isotropic, so the slowest modes supported on the island behave as 
Laplacian eigenfunctions on a bounded domain of size $\alpha_p$, yielding the scaling
\begin{equation}
\gamma_k(\psi^p) \sim D\,\alpha_p^{-2} k^2.
\end{equation}
Differences in $\alpha_p$ determine where modes from different island families appear in 
the global spectrum.

\subsection*{Core curvature and advective modes}

Near the elliptic fixed point, $H_{\mathrm{eff}}$ admits a quadratic expansion
\begin{equation}
H_{\mathrm{eff}}(I) \;=\; H_0 + \tfrac{1}{2}\beta_p I^2 + O(I^3),
\end{equation}
with $\beta_p = d\Omega/dI|_{I=0}$. This curvature fixes the local oscillator frequency 
and hence the spacing of the advective ladder $\phi^p_{q,k,m}$. Diffusion introduces a 
transverse length scale $O(\sqrt{D})$ in action space, so that the oscillator spectrum 
obeys
\begin{equation}
\gamma_k(\phi^p) \sim \sqrt{D}\,\beta_p^{1/2} k.
\end{equation}

\subsection*{Spectral phase and advective branches}

For an eigenvalue $\lambda = |\lambda| e^{\pm i\theta}$ of the one-period 
advection-diffusion operator, the argument $\theta$ encodes the rotation frequency of the 
underlying Koopman mode. Near an elliptic fixed point, the effective Hamiltonian provides 
a local rotation frequency $\Omega_0 \approx \Omega(I)|_{I=0}$, which defines a 
fundamental advective phase
\begin{equation}
\theta_{\mathrm{adv}} = \Omega_0.
\end{equation}

Advective modes are organized by azimuthal wavenumber $q$, corresponding to the number of 
windings around the elliptic core. Modes with azimuthal index $q$ rotate $q$ times per 
iterate, producing spectral phases
\begin{equation}
\theta_q = q\,\theta_{\mathrm{adv}} \pmod{2\pi}.
\end{equation}
In the raw spectra (Fig.~\ref{spectra}), this appears as families of eigenvalues aligned 
along discrete rays at $\theta = q\,\theta_{\mathrm{adv}}$, forming the vertical 
alignment bands visible in the spectral plots. The radial index $k$ within each azimuthal 
family labels successive radial excitations at fixed $q$, corresponding to progressively 
larger action levels $I_k$ and additional radial nodes in the eigenfunction.

As $D$ decreases, more radial levels become resolved within each azimuthal sector, and 
the discrete harmonic structure $\{q\,\theta_{\mathrm{adv}}\}$ becomes increasingly 
apparent. In the limit $D\to0$, the spectrum approaches a continuous arc on the unit 
circle, consistent with the continuous Koopman spectrum of the diffusion-free map, but 
retains the discrete harmonic organization imposed by the elliptic core geometry.

\end{document}